\documentclass[fleqn,usenatbib]{mnras}

\usepackage{newtxtext,newtxmath}

\usepackage[T1]{fontenc}
\usepackage{ae,aecompl}
\usepackage{soul}

\usepackage{graphicx}	
\usepackage{amsmath}	
\usepackage[usenames]{xcolor}
\usepackage{natbib}
\usepackage{hyperref}
\usepackage{xfrac}
\usepackage{cases}

\def\equationautorefname~#1\null{equation~(#1)}

\DeclareMathAlphabet{\mathpzc}{OT1}{pzc}{m}{it}\definecolor{purple}{RGB}{160,32,240}
\newcommand{\nick}[1]{\textcolor{black}{#1}}
\newcommand{\rev}[1]{\textcolor{black}{#1}}
\newcommand{\athena}{\texttt{Athena++}}
\newcommand{\pgn}{\texttt{PEnGUIn}}

\newcommand{\Mstar}{M_{\star}}
\newcommand{\Mearth}{M_{\oplus}}
\newcommand{\Mj}{M_{\rm J}}

\newcommand{\Sigg}{\Sigma_{\rm g}}
\newcommand{\rhog}{\rho_{\rm g}}

\newcommand{\gcm}{\rm g/cm^{2}}
\newcommand{\rb}{r_{\rm Bondi}}
\newcommand{\Rp}{R_{\rm p}}
\newcommand{\Mp}{M_{\rm p}}
\newcommand{\rh}{r_{\rm Hill}}
\newcommand{\qth}{q_{\rm th}}
\newcommand{\Hp}{H_{\rm p}}
\newcommand{\mdot}{\dot{M}_{\rm p}}
\newcommand{\mdotb}{\dot{M}_{\rm Bondi} }

\newcommand{\mdottwo}{\dot{M}_{\rm Hill,\,2D} }
\newcommand{\mdotthree}{\dot{M}_{\rm Hill,\,3D} }
\newcommand{\mdotin}{\dot{M}_{\rm p,in}}
\newcommand{\mdotout}{\dot{M}_{\rm p,out}}

\newcommand{\cs}{c_{\rm s}}
\newcommand{\mtd}{\mathrm{min}\left(t_{\rm double}\right)}
\newcommand{\Omegap}{\Omega_{\rm p}}
\newcommand{\minrh}{\mathrm{min}(r_{\rm Bondi},\,r_{\rm Hill})}
\newcommand{\minh}{\mathrm{min}(r_{\rm Bondi},\,H_{\rm p})}

\title[Protoplanet accretion rates]{The maximum accretion rate of a protoplanet: how fast can runaway be?}

\author[Choksi et al.]{Nick Choksi$^{1 \href{https://orcid.org/0000-0003-0690-1056}{\includegraphics[scale=0.4]{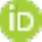}}}$\thanks{E-mail: nchoksi@berkeley.edu}, Eugene Chiang$^{1,2}$, Jeffrey Fung$^{3}$, and Zhaohuan Zhu$^{4,5}$ \\ 
$^{1}$Department of Astronomy, Theoretical Astrophysics Center, and Center for Integrative Planetary Science, University of California, Berkeley, CA 94720, USA\\
$^{2}$Department of Earth and Planetary Science, University of California, Berkeley, CA 94720, USA \\
$^{3}$Department of Physics and Astronomy, Clemson University, Clemson, SC 29634, USA \\
$^{4}$Department of Physics and Astronomy, University of Nevada, Las Vegas, NV 89154, USA \\
$^{5}$Nevada Center for Astrophysics, University of Nevada, Las Vegas, NV 89154, USA \\
}

\date{Released \today}
\pubyear{2023}

\begin{document}
\label{firstpage}
\pagerange{\pageref{firstpage}--\pageref{lastpage}}
\maketitle

\begin{abstract} 
The hunt is on for dozens of protoplanets hypothesised to reside in protoplanetary discs with imaged gaps. How bright these planets are, and what they will grow to become, depend on their accretion rates, which may be in the runaway regime. Using 3D global simulations we calculate maximum gas accretion rates for planet masses $\Mp$ from 1$\,\Mearth$ to $10\,\Mj$. 
When the planet is small enough that its sphere of influence is fully embedded in the disc, with a Bondi radius $\rb$ smaller than the disc's scale height $\Hp$ --- such planets have thermal mass parameters $\qth \equiv (\Mp/\Mstar) / (\Hp/\Rp)^3 \lesssim 0.3$, for host stellar mass $\Mstar$ and orbital radius $\Rp$ --- the maximum accretion rate follows a Bondi scaling, with $\max \dot{M}_{\rm p} \propto \rhog \Mp^2 / (\Hp/\Rp)^3$ for ambient disc density $\rhog$. 
For more massive planets with 
$0.3 \lesssim \qth \lesssim 10$, the Hill sphere replaces the Bondi sphere as the gravitational sphere of influence, and $\max \dot{M}_{\rm p} \propto \rhog \Mp^1$, with no dependence on $\Hp/\Rp$. In the strongly superthermal limit when $\qth \gtrsim 10$, the Hill sphere pops well out of the disc, and $\max \dot{M}_{\rm p} \propto \rhog M_{\rm p}^{2/3} (\Hp/\Rp)^1$. Applied to the two confirmed protoplanets PDS 70b and c, our numerically calibrated maximum accretion rates imply their Jupiter-like masses may increase by up to a factor of $\sim$2 before their parent disc dissipates.
\end{abstract}



\section{Introduction}
\label{sec:Intro}

The Atacama Large Millimeter Array (ALMA) is imaging circumstellar discs at high angular resolution and finding annular gaps in dust \citep[][]{alma_etal_2015, huang_etal_2018, cieza_etal_2019} and gas \citep{isella_etal_2016, fedele_etal_2017, favre_etal_2019, zhang_etal_2021}. A popular interpretation is that these gaps are opened by embedded planets and the  density waves they excite 
(\citealt{goldreich_tremaine_1980, goodman_rafikov_2001, kanagawa_etal_2016, zhang_etal_2018, dong_fung_2017, bae_etal_2017}). Velocity-resolved channel maps of gas emission lines also reveal non-Keplerian gas motions that could be stirred by  planets \citep[][]{teague_etal_2018c, teague_etal_2019_nature, pinte_etal_2020, pinte_etal_2023}.
Dozens of potential planets have been identified; see Table \ref{tab:data} for a compilation. Efforts to confirm their presence by direct imaging are accelerating  \citep{cugno_etal_2019, zurlo_etal_2020, asensio-torres_etal_2021, jorquera_etal_2021, facchini_etal_2021, huelamo_etal_2022, currie_etal_2022,follette_etal_2022, cugno_etal_2023}, but so far only the protoplanets PDS 70b and c have been captured in their own light \citep{haffert_etal_2019, wang_etal_2020, wang_etal_2021, zhou_etal_2021}.

Prospects for direct imaging depend critically on accretion luminosities. The planet masses $\Mp$ inferred from fitting disc substructures are usually $\gtrsim 10\,\Mearth$
(\citealt{zhang_etal_2018}; also our Table \ref{tab:data}), \nick{large enough that
the planets may have acquired massive gas envelopes \citep[e.g.][]{piso_youdin_2014}. 
} The self-gravity of these envelopes can lead to ``runaway'' accretion whereby the mass doubling time of a planet $M_{\rm p}/\dot{M}_{\rm p}$ decreases with increasing $M_{\rm p}$ \citep[e.g.][]{pollack_etal_1996}. \nick{Runaway can be thermodynamic, brought about by large envelope luminosities and short cooling times in quasi-hydrostatic equilibrium, or hydrodynamic, characterized by flows that accelerate to planetary free-fall velocities \citep{mizuno_etal_1978, ginzburg_chiang_2019a}.} 

\rev{The outcome of runaway is commonly presumed to be Jupiter-sized gas giants, though how this process unfolds and in particular how it ends remain uncertain.  
What are the relevant planet accretion rates, and how do they depend on planet mass and disc parameters?}
Numerical simulations have provided data and fitting formulae in various patches of parameter space \citep[e.g.][]{tanigawa_watanabe_2002, dangelo_etal_2003, machida_etal_2010, bethune_rafikov_2019}, but we are not aware of an analytic or unifying theory. To the usual problems associated
with accretion --- how material cools and how it sheds angular momentum
--- we need to add, for a protoplanet orbiting a star, how
gas moves in their combined potential, including rotational forces, in 3D. 
\citet{lambrechts_etal_2019b} point out that what several large-scale disc-planet simulations report as mass accretion rates are actually only upper limits, as permanent accretion of mass depends on smaller-scale physics (e.g.~cooling of the planetary interior) which simulations typically do not resolve. 

In trying to understand from first principles how protoplanets accrete, \citet{ginzburg_chiang_2019a} started with the simplest model,
  that runaway accretion takes the form of Bondi accretion from a uniform medium with no angular momentum (see, e.g., the textbook by \citealt{frank_king_raine}).
The assumption of uniform background density would be justified if the planet were fully embedded in the disc, i.e. if its gravitational radius of influence, measured by the Bondi radius $\rb$, were smaller than the local circumstellar disc height $\Hp$. The ratio of the two lengths is the thermal mass parameter
\begin{align}
    \qth &\equiv \frac{\rb}{\Hp} \nonumber \\ 
        &= \frac{\Mp }{\Mstar (\Hp/\Rp)^3} \,,
    \label{eqn:qth}
\end{align}
where $\rb = G\Mp/\cs^2$, $G$ is the gravitational constant, $\Mp$ is the planet mass, $\Mstar$ is the host stellar mass, $\Hp = \cs/\Omegap$, and $\Omegap$ is the planet's Keplerian frequency at orbital radius $\Rp$. On the one hand, roughly half of hypothesised gap-opening planets have $\qth \lesssim 1$ (see Table \ref{tab:data}), motivating a Bondi-like accretion rate that scales as $\dot{M}_{\rm p} \propto \Mp^2$. On the other hand, the spherically symmetric Bondi solution ignores the meridional flow patterns seen in 3D simulations \citep[][]{szulagyi_etal_2014, fung_etal_2015, ormel_etal_2015}. 


More massive ``superthermal'' planets with $\qth \gtrsim 1$ sample more of the disc's
vertical density gradient. Stellar tidal forces also enter; these
pare accreting material down to the planet's Hill sphere, which in the
superthermal regime now lies inside the Bondi radius.
As with subthermal planets, there seems no consensus for how the
superthermal accretion
rate scales with input parameters.
A simple argument based on the Hill sphere and Keplerian shear yields an accretion rate $\mdot \propto
\Mp^{2/3}$ (e.g.~\citealt{rosenthal_etal_2020}, their equation 7, and
references therein). But many studies
(e.g. \citealt{mordasini_etal_2015, elee_2019, lambrechts_etal_2019b})
adopt the empirical scaling $\dot{M}_{\rm p} \propto \Mp^{4/3}$ 
reported by \cite{tanigawa_watanabe_2002} from their 2D numerical
simulations. The two options lie on opposite sides of the
$\dot{M}_{\rm p} \propto \Mp^1$ scaling which divides power-law
growth from super-exponential runaway growth.

Our goal here is to help 
clear up what seems like a longstanding confusion. 
We utilize 3D isothermal numerical simulations of planet-disk interactions, similar to those used by others, to decide how the protoplanet accretion rate $\dot{M}_{\rm p}$ depends on planet mass $\Mp$, local disc gas density $\rho_{\rm g}$, and disc aspect ratio $\Hp/\Rp$, starting in the subthermal regime ($\sim$1 $M_\oplus$) and working our way systematically to the superthermal limit ($\sim$10 $M_{\rm J}$). Actually our findings will be restricted to $\max \dot{M}_{\rm p}$, as we track only how much mass potentially accretes upon entering a planet's gravitational sphere of influence, not how much actually accretes (see also \citealt{lambrechts_etal_2019b}). Section \ref{sec:numerics} details our numerical methods. Section \ref{sec:results} reports $\max \dot{M}_{\rm p}$ and how its dependence on input parameters can be understood and reproduced using simple arguments. Section \ref{sec:summary_discussion} summarises, discusses how our work makes sense of previous numerical studies, and connects to observations.

\begin{table*}
\begin{tabular}{cccccccccc}
\hline 
\\[-2mm]
(1) & (2) & (3) & (4) & (5) & (6) & (7) & (8) & (9) & (10)  \\ 
Name & $M_{\star} [M_\odot]$ & $t_{\rm age}\,\,[\rm Myr]$ & $R_{\rm p}\,\,[\rm au]$ & $M_{\rm p}\,\,[M_{\rm J}]$ & $\Sigma_{\rm g}\,\,[\rm g/cm^{2}]$ & $H_{\rm p}/R_{\rm p}$ & $q_{\rm th}$ & $\log_{10}\left(\frac{\dot{M}_{\rm p,in}}{M_{\rm J}/{\rm Myr}}\right)$ & $\log_{10}\left[ \frac{\min (t_{\rm double})}{t_{\rm age}}\right]$ \\
\hline 
\\[-2mm]
Sz 114 & 0.17 & $1$ & 39 & 0.01-0.02 & ... & $0.1$ & 0.06-0.1 & ... & ... \\
GW Lup & 0.46 & $2$ & 74 & 0.007-0.03 & ... & $0.08$ & 0.03-0.1 & ... & ... \\
Elias 20 & 0.48 & $0.8$ & 25 & 0.03-0.07 & ... & $0.08$ & 0.1-0.3 & ... & ... \\
Elias 27 & 0.49 & $0.8$ & 69 & 0.01-0.07 & ... & $0.09$ & 0.03-0.2 & ... & ... \\
RU Lup & 0.63 & $0.5$ & 29 & 0.03-0.07 & ... & $0.07$ & 0.2-0.3 & ... & ... \\
SR 4 & 0.68 & $0.8$ & 11 & 0.2-2 & ... & $0.05$ & 2-30 & ... & ... \\
Elias 24 & 0.8 & $2$ & 55 & 0.5-5 & ... & $0.09$ & 0.9-9 & ... & ... \\
TW Hya-G1 & 0.8 & $8$ & 21 & 0.03-0.3 & 0.04-3 & $0.08$ & 0.07-0.7 & ${-2}, {1}$ & ${-3}, {-1}$ \\
TW Hya-G2 & 0.8 & $8$ & 85 & 0.02-0.2 & 0.008-0.2 & $0.09$ & 0.03-0.3 & ${-3}, {1}$ & ${-2}, {0}$ \\
Sz 129 & 0.83 & $4$ & 41 & 0.02-0.03 & ... & $0.06$ & 0.09-0.2 & ... & ... \\
DoAr 25-G1 & 0.95 & $2$ & 98 & 0.07-0.1 & ... & $0.07$ & 0.2-0.3 & ... & ... \\
DoAr 25-G2 & 0.95 & $2$ & 125 & 0.02-0.03 & ... & $0.07$ & 0.05-0.1 & ... & ... \\
IM Lup & 1.1 & $0.5$ & 117 & 0.03-0.1 & 0.1-10 & $0.1$ & 0.03-0.09 & ${-1}, {2}$ & ${-2}, {0}$ \\
AS 209-G1 & 1.2 & $1$ & 9 & 0.2-2 & ... & $0.04$ & 3-30 & ... & ... \\
AS 209-G2 & 1.2 & $1$ & 99 & 0.1-0.7 & 0.04-0.4 & $0.06$ & 0.4-3 & ${0}, {1}$ & ${-2}, {-1}$ \\
$\rm AS\,209{\text -}G3^{\ast}$ & 1.2 & $1$ & 240 & 0.01-0.05 & 0.0002-0.1 & $0.07$ & 0.02-0.1 & ${-4}, {0}$ & ${-1}, {3}$ \\
HD 142666 & 1.58 & $10$ & 16 & 0.03-0.3 & ... & $0.05$ & 0.2-2 & ... & ... \\
HD 169142 & 1.65 & $10$ & 37 & 0.1-1 & 0.1-0.2 & $0.07$ & 0.2-2 & ${0}, {1}$ & ${-2}, {-2}$ \\
HD 143006-G1 & 1.78 & $4$ & 22 & 1-20 & ... & $0.04$ & 10-200 & ... & ... \\
HD 163296-G1 & 2.0 & $10$ & 10 & 0.1-0.7 & ... & $0.07$ & 0.1-1 & ... & ... \\
HD 163296-G2 & 2.0 & $10$ & 48 & 0.3-2 & 1-40 & $0.08$ & 0.3-2 & ${1}, {4}$ & ${-4}, {-3}$ \\
HD 163296-G3 & 2.0 & $10$ & 86 & 0.03-1 & 0.1-20 & $0.08$ & 0.03-1 & ${-2}, {3}$ & ${-4}, {-1}$ \\
$\rm HD\,163296{\text -}G4^{\ast}$ & 2.0 & $10$ & 137 & 0.002-1 & 0.2-7 & $0.09$ & 0.002-1 & ${-3}, {3}$ & ${-4}, {0}$ \\
$\rm HD\,163296{\text -}G234alt^{\ast}$ & 2.0 & $10$ & 108 & 0.2 & 0.5-10 & $0.08$ & 0.2 & ${1}, {2}$ & ${-4}, {-2}$ \\
$\rm HD\,163296{\text -}G5^{\ast}$ & 2.0 & $10$ & 260 & 0.01-2 & 0.1-2 & $0.09$ & 0.009-1 & ${-2}, {2}$ & ${-3}, {-1}$ \\
PDS 70b & 1.0 & $5$ & 22 & 1-10 & 0.0008-0.08 & $0.07$ & 3-30 & ${-1}, {1}$ & ${-1}, {1}$ \\
PDS 70c & 1.0 & $5$ & 34 & 1-10 & 0.0008-0.08 & $0.08$ & 2-20 & ${-1}, {1}$ & ${-1}, {1}$ \\
\hline 
\end{tabular}
\caption{Properties of gapped discs and the planets hypothesized (confirmed in the case of PDS 70) to reside within them, adapted from \protect \cite{choksi_chiang_2022}. Column headings: (1) System name. We append ``G\#'' to distinguish between different gaps in a given system. Asterisks mark new entries not tabulated by \protect \cite{choksi_chiang_2022} and are further described in Appendix \ref{sec:data}. The entry ``HD 163296-G234alt'' refers to the possibility that the gaps at 48, 86, and 137 au in HD 163296 do not contain planets but are opened by a single planet at 108 au \protect \citep{dong_etal_2018}. (2) Stellar mass (3) Stellar age (4) Planet orbital radius. For HD 169142, we adjusted $\Rp$ to 37 au to match the location of an unconfirmed compact source in the gap \citep{hammond_etal_2023}. (5) Planet mass. In most cases $\Mp$ is estimated from the width of the gap (\citealt{zhang_etal_2018}), assuming the Shakura-Sunyaev viscosity parameter $\alpha = 10^{-5} - 10^{-3}$.
(6) Gas surface density in the gap, based on  spatially resolved C$^{18}$O emission. The quoted range accounts for uncertainty in the CO:H$_2$ abundance.
(7) Disc aspect ratio at the planet's position, estimated either by assuming the disc is passively heated by its host star or from fits to mm-wave observations. (8) Planet thermal mass parameter $\qth = (\Mp/\Mstar)/ (\Hp/\Rp)^3$. (9) \rev{Maximum planetary accretion rate $\mdotin$, equal to the rate at which gas from the nebula flows into the Bondi or Hill sphere, whichever is smaller. Commas separate minimum and maximum sink-cell accretion rates obtained by inserting the range of possible values for $\qth$, $\Hp/\Rp$, and $\rhog = \Sigg/\left(\sqrt{2\pi} \Hp\right)$ into equations \ref{eqn:sum1}-\ref{eqn:sum3}. Values are rounded to the nearest order of magnitude.} (10) Lower bound on the mass doubling time, $\min \, (t_{\rm double}) = \Mp/\mdotin$, divided by the system age. Commas separate minimum and maximum values rounded to the nearest order of magnitude and are plotted in Fig.~\ref{fig:tdouble}.}
\label{tab:data}
\end{table*}

\section{Simulation setup}
\label{sec:numerics}
Most of our simulations are performed with the Eulerian hydrodynamics code $\athena$ \citep{stone_etal_2020}, outfitted with a second-order van Leer time integrator (\texttt{integrator = vl2}), a second-order piecewise linear spatial reconstruction of the fluid variables (\texttt{xorder = 2}), and the Harten-Lax-van Leer-Einfeldt Riemann solver (\texttt{--flux hlle}).
For some regions of parameter space, we check our results against published simulations by \citet{fung_etal_2019} that used the 
Lagrangian-remap, GPU code $\pgn$ \citep[][]{fung_etal_2015}. The setup of our $\athena$ simulations is described below, with differences between $\pgn$ and $\athena$ highlighted.

\subsection{Equations solved}\label{subsec:eqsolve}
$\athena$ solves the 3D Euler equations: 
\begin{align}
&\frac{\partial \rho}{\partial t} + \nabla \cdot \left( \rho \mathbf{v} \right) = 0 
\label{eqn:continuity} \\  
&\frac{\partial \left( \rho \mathbf{v} \right)}{\partial t} +  \nabla \cdot \left(  \rho \mathbf{v} \otimes \mathbf{v} \right)   = -\nabla P - \rho \nabla \Phi
\label{eqn:euler}
\end{align}
where $\rho$, $\mathbf{v}$, and $P$ are the gas density, velocity, and pressure, and $\Phi$ is the gravitational potential. We use an isothermal equation of state
\begin{equation}
P = \rho \cs^2
\label{eqn:eos}
\end{equation}
with constant sound speed $\cs$. 
In the hydrodynamic runaway 
phase of giant planet formation, the planet's atmosphere cools rapidly and so the isothermal approximation seems appropriate, at least on Bondi sphere scales \citep{piso_youdin_2014, elee_chiang_2015, ginzburg_chiang_2019a}.

Simulations are performed in the frame rotating at the planet's orbital angular frequency $\Omegap = 1$, using spherical coordinates $(R,\,\Theta,\,\Psi)$ centred on the star, where $R$ is radius, and $\Theta$ and $\Psi$ are the polar and azimuthal angles, respectively. In this frame the planet is fixed at $(\Rp,\,\Theta_{\rm p},\,\Psi_{\rm p}) =  (1,\,\pi/2,\,\pi)$.

The gravitational potential is the sum of the potentials due to the star of mass $M_\star$ and the planet of mass $\Mp$, plus the indirect potential arising from our 
star-centred grid:
\begin{align}
\Phi &= -\frac{G\Mstar}{R} -  \frac{G\Mp}{\sqrt{R^2 + \Rp^2 - R\Rp\sin\Theta \cos (\Psi - \Psi_{\rm p} ) }} \times f_{\rm soft} \nonumber \\
&+ \frac{G\Mp R\sin\Theta \cos (\Psi - \Psi_{\rm p} )}{\Rp^2}  
\label{eqn:potential}
\end{align}
where $G$ is the gravitational constant. When the distance from the planet $d = \sqrt{R^2 + \Rp^2 - R\Rp\sin\Theta \cos(\Psi - \Psi_{\rm p})}$ exceeds $r_{\rm soft}$, we set $f_{\rm soft} = 1$. Closer to the planet, the potential is softened ($f_{\rm soft} < 1$) according to
\begin{align}
f_{\rm soft} &= \left(\frac{d}{r_{\rm soft} }\right)^4 - 2\left(\frac{d}{r_{\rm soft}}\right)^3 + 2\left(\frac{d}{r_{\rm soft}}\right)  \hspace{1cm} \,\,\mathrm{if}\,\,d < r_{\rm soft} \,.
\end{align}
We set $r_{\rm soft}$ to three times the smallest cell size. The $\pgn$ simulations use a different softening prescription given by equation 11 of \cite{fung_etal_2019}. 

\rev{A subset of our $\athena$ runs simulate planetary accretion using sink cells.
Gas densities inside cells for which $d < r_{\rm sink}$ are depleted at a rate 
\begin{equation}
\frac{\partial \rho}{\partial t} = -\frac{\rho}{\tau_{\rm sink}} 
\end{equation}
where $r_{\rm sink} = \minrh/10$, $\rb = G\Mp/\cs^2$, $\rh = 3^{-1/3}\left(\Mp/\Mstar\right)^{1/3}\Rp$, and $\tau_{\rm sink} = r_{\rm sink}/\cs$. At our fiducial resolution, $r_{\rm sink} \simeq 2 r_{\rm soft} = 0.1 r_{\rm Bondi}$ 
for subthermal runs. 
For superthermal runs, $r_{\rm sink} \simeq r_{\rm soft} = 0.1 r_{\rm Hill}$. The mass removed is not added to the planet; for typical parameters of non-self-gravitating discs, the mass removed over the simulation duration is $\ll \Mp$. In Appendix \ref{sec:convergence} we test the sensivitity of our results to $r_{\rm sink}$.}

\subsection{Initial and boundary conditions}
\label{subsubsec:ics_bcs}
In the $\athena$ runs, the planet mass is initially zero and is ramped up to its final mass $M_{\rm p,final}$ over one orbital period $2\pi/\Omegap$:
\begin{alignat}{2}
M_{\rm p}(t) &= M_{\rm p,final}\sin^2 \left [ \frac{t}{2\pi \Omegap^{-1}} \times  \frac{\pi}{2} \right] \,\,\,\, &&\mathrm{for}\,\,\,\,  t < 2\pi\Omegap^{-1} \nonumber \\ 
 &= M_{\rm p,final}\,\, &&\mathrm{for}\,\, \,\, t \geq 2\pi\Omegap^{-1}.
\end{alignat}
\rev{(In those runs that use sink cells, the sink-cell prescription is always applied, including during this initial ramp up.)} In the $\pgn$ simulations the planet mass is not ramped up.

We assume the disc is initially axisymmetric with a density profile
\begin{align}
\rho(R,\,\Theta) = \rho_0\left(\frac{R\sin\Theta}{\Rp} \right)^{-\alpha} \exp\left[-\frac{G\Mstar}{R\cs^2}\left(\frac{1}{\sin\Theta} - 1 \right) \right].
\label{eqn:rhoinit}
\end{align}
Here $\rho_0$ is the initial midplane gas density at the planet's position. Since we ignore gas self-gravity, we are free to take $\rho_0 = 1$. The $\athena$ simulations use $\alpha = 1.5$ and the $\pgn$ simulations use $\alpha = 3$. Because the planet is fed by co-orbital material, the value of $\alpha$ should have little impact on our results for accretion rate. When $\pi/2 - \Theta \ll 1$, equation \ref{eqn:rhoinit} is a Gaussian in the vertical direction with scale height $H = \cs/\sqrt{GM_\star/R^3}$. At the planet's position, $H = \Hp = \cs/\Omegap$. 

The initial velocity field of the gas is purely azimuthal and constant on cylinders:
\begin{align}
v_{\Psi} = \sqrt{\frac{G\Mstar}{R\sin\Theta} - \alpha \cs^2} \,.
\end{align}
The second term involving $\cs$ accounts for how the disc's radial pressure gradient slows rotation. 

We define the planet's Bondi radius as $\rb \equiv G\Mp/\cs^2$ and the thermal mass parameter as $\qth \equiv \rb/\Hp$. For subthermal planets $0.02 \leq \qth < 1$, 
the simulation domain spans $R = \Rp \pm 10 \rb$, $\Psi = \Psi_{\rm p} \pm 10 \rb/\Rp$, and $\Theta = \pi/2$ to $\pi/2 - 3\Hp/\Rp$. Only the upper half of the disc at $\Theta < \pi/2$ is simulated; the flow is assumed symmetric about the midplane, with boundary conditions there as appropriate (e.g., $v_\Theta = 0$ at $\Theta = \pi/2$). Runs with smaller $\qth$ are especially computationally costly, so for $\qth \leq 0.05$ we limit the upper boundary to $\Theta = \pi/2 - 30\rb/\Rp$. At all boundaries except for the midplane the flow is fixed to its initial conditions. For subthermal runs in $\pgn$, the simulation domain and boundary conditions are the same as in $\athena$, except in $\pgn$ the full 2$\pi$ in azimuth is simulated with periodic boundary conditions, and a reflecting boundary condition is used for the $\Theta$-boundary above the midplane. For superthermal runs where $\qth \geq 1$, both $\athena$ and $\pgn$ use radial domains that span $\pm 10\Hp$ around the planet and azimuthal domains that cover 2$\pi$.

Wave-killing zones in $\athena$ damp reflections near the radial boundaries:
\begin{alignat}{2} 
\frac{\partial X}{\partial t} &= -\left(\frac{X - X(t=0)}{\tau} \right)K(R) \nonumber \\ 
K(R) &= 1 -  \sin^2\left[ \frac{\pi}{2}  \left( \frac{R - R_1}  { R_{\rm kill,1} - R_1 } \right)   \right] \,\,&&\mathrm{if}\,\,R < R_{\rm kill,1} \nonumber  \\ 
 &=  \sin^2\left[ \frac{\pi}{2} \left( \frac{R - R_{\rm kill,2}}  {R_2 -  R_{\rm kill,2}  } \right)\right]  \,\,&& \mathrm{if}\,\,R > R_{\rm kill,2}  \,, 
\end{alignat}
where $X$ is either mass density or momentum density and $\tau$ is a damping timescale that we set to $0.1 \times 2 \pi \Omegap^{-1}$. The inner and outer radial boundaries of the simulation domain are $R_1$ and $R_2$. We place the damping boundaries $R_{\rm kill,1}$ and $R_{\rm kill,2}$ so that the two zones encompass the inner and outer 10\% of the radial domain, respectively. The $\pgn$ simulations use a wave-killing prescription given by equation 16 of \cite{fung_etal_2019}.

\subsection{Resolution}
\label{subsubsec:refinement}

We use static mesh refinement in $\athena$. The highest resolution region is approximately a sphere of radius three times $x = \minh$ centred on the planet, having boundaries
\begin{align}
    \left(R_{\rm min}, R_{\rm max} \right) &= \left(\Rp - 3x,    \Rp + 3x\right ) \nonumber \\ 
    \left(\Theta_{\rm min}, \Theta_{\rm max}\right) &= \left(\pi/2 - 3x/\Rp,\,\pi/2 \right) \nonumber  \\
        \left(\Psi_{\rm min}, \Psi_{\rm max}\right ) &= \left(\Psi_{\rm p} - 3x/\Rp,   \Psi_{\rm p} + 3x/\Rp \right ) \,.
    \label{eqn:smr}
\end{align}
The cells in this region have width $\Delta R = \Rp \Delta \Theta = \Rp \Delta \Psi = x/64$ in subthermal runs and $x/32$ in superthermal runs. Outside of this region, cell widths increase by successive factors of two until they reach $\Delta R = \Rp \Delta \Theta = \Rp \Delta \Psi = x/8$. We 
test the convergence of our results with resolution in Appendix \ref{sec:convergence}.

The $\pgn$ simulations also boost resolution near the planet. Instead of using discrete levels of refinement as in $\athena$, $\pgn$ smoothly changes the cell widths as prescribed in section 2.1.2 of \cite{fung_etal_2019}. The cell width at the planet's position in $\pgn$ is $\Delta R = \Rp \Delta \Theta = \Rp \Delta \Psi = x/64$. 
\citet[][their fig.~1]{fung_etal_2019} show that their results for $x/64$ using $\pgn$ converge to within a few percent of their results for $x/128$ at distances $\gtrsim 0.1 \rb$ from the planet.

\subsection{Run duration and steady state}

\rev{Simulations with $\athena$ are run for at least $15 \Omegap^{-1}$, long enough that over much of our parameter space, a quasi-steady state is reached in the flow patterns around the planet. The $\pgn$ simulations are run for nearly $10\times$ longer, and as we show below, yield results consistent with our $\athena$ runs (see also section 2.1 of \citealt{fung_etal_2019} which notes that near-steady states are reached after $\sim$2 orbits). A handful of $\athena$ runs are extended out to hundreds of $\Omegap^{-1}$ and evince no change in behaviour from our standard runs.}

\rev{Our aim in this paper is to understand planetary flow patterns on dynamical timescales, i.e., on sound-crossing timescales of $\rb/\cs$ or local shearing timescales. These are of order $\Omegap^{-1}$ or shorter. Thus our finding that steady states are achieved after just a few orbits is not surprising. Over longer timescales, and for the most massive planets simulated, we see annular gaps gradually open in the planet's co-orbital region. We show in section \ref{subsec:gaps} that our results can be straightforwardly scaled by the time-evolving disc density in these runs.}

\section{Results}
\label{sec:results}
Although our simulations are performed in spherical coordinates $(R,\,\Theta, \Psi)$ centred on the star, in analysing our results we will use spherical coordinates $(r,\,\theta,\,\phi)$ and cylindrical coordinates $(r_{\rm cyl},\,z,\,\phi)$ 
centred on the planet. \nick{We use nearest-neighbor interpolation to calculate fluid properties between cell centres.} 

Our focus in this paper is on $\mdotin$, defined as the mass per time entering a sphere of given radius $r$ centred on the planet. It is a ``one-way'' rate because it counts only the mass whose radial velocity $v_r < 0$. The analogous outflow rate $\mdotout$ counts only the mass whose $v_r > 0$. By construction both $\mdotin$ and $\mdotout$ are positive; the net mass accretion rate onto the planet is $\dot{M}_{\rm p} = \mdotin - \mdotout$. 

\rev{We interpret our results for $\mdotin$, obtained both with and without sink cells (section \ref{subsec:eqsolve}), as upper limits on the true mass accretion rate $\mdot$. We anticipate that $\mdotin$ measured with sink cells will be at least as large as $\mdotin$ measured without, and confirm this below. Actually we will find that the two cases yield rather similar results. Our measurements of $\mdotin$ should be robust insofar as  the inflow is supersonic and therefore independent of downstream boundary conditions.\footnote{In our isothermal simulations the inflow along the planet's polar axis is supersonic. If in reality the inflow were adiabatic and subsonic (\citealt{fung_etal_2019}), we would expect $\mdotin$ to be lower. This paper's measurements of $\mdotin$ under isothermal conditions would still stand as hard upper limits on the true $\dot{M}_{\rm p}$.} This robustness will be evidenced by the similarity between our results for $\mdotin$ without sink cells (sections \ref{subsec:subthermal}-\ref{subsec:gaps}) and with them (section \ref{subsec:sink}).}

\begin{figure} 
\includegraphics[width=\columnwidth]{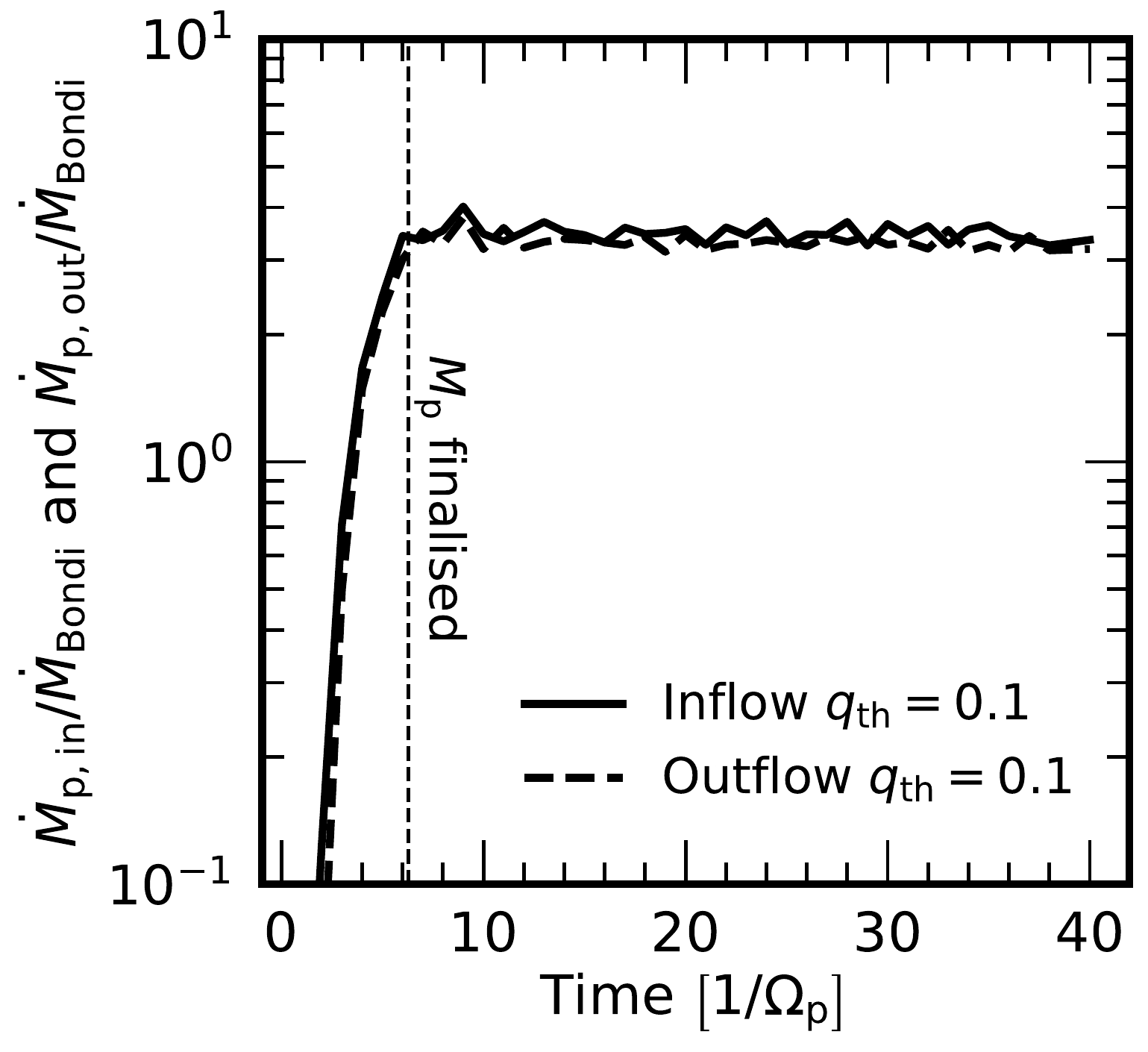}
\caption{
\rev{Time evolution of the inflow rate (solid curve) and outflow rate (dashed) evaluated at $r = \rb$ for our $\qth = 0.1$, $\Hp/\Rp = 0.035$ $\athena$ simulation without a sink cell.} The planet mass $\Mp$ is ramped up from 0 at $t = 0$ to its final value at $t = 2\pi \Omegap^{-1}$ (vertical line). Beyond this time, the simulation is in a quasi-steady state where outflow nearly balances inflow. In reality the difference between inflow and outflow --- i.e. the true net accretion rate --- depends on the circumplanetary physics of cooling and viscosity which our simulations do not capture. Thus our paper focuses on just the inflow rate as an upper limit on the true accretion rate.}
  \label{fig:Mdot_times}
\end{figure}

By contrast our simulated outflow rates $\mdotout$, and by extension the net rates $\dot{M}_{\rm p}$, are problematic to interpret. Although physically some outflow is expected because a fraction of the inflowing material may have too much energy to become bound to the planet, or too much angular momentum to cross the centrifugal barrier, exactly what this fraction is cannot be determined without accounting for cooling and viscosity (see also \citealt{lambrechts_etal_2019b}). \rev{In lieu of incorporating this circumplanetary physics, our simulations (and those of many others) 
use softened gravitational potentials, with or without sink cells. With a sink cell, we expect $\mdotout \ll \mdotin$.
Without a sink cell, our simulations settle into a quasi-steady state in which $\mdotout$ balances $\mdotin$, as illustrated in Figure \ref{fig:Mdot_times}. The balance is good to within $\sim$15\% in the $\athena$ simulations, and a few percent in the $\pgn$ simulations (\citealt{fung_etal_2019}, their figure 17). Whether or not we use a sink cell, in all of these oversimplified numerical treatments, $\mdotout$ lacks physical meaning (cf.~\citealt{ormel_etal_2015}). Accordingly, we concentrate  on $\mdotin$ and understanding its physical dependence on parameters.}

Because we simulate only half the disc and assume symmetry about the midplane, mass flow rates reported in this paper are 2$\times$ those simulated.

\subsection{Subthermal limit}
\label{subsec:subthermal}

\begin{figure} 
\includegraphics[width=\columnwidth]{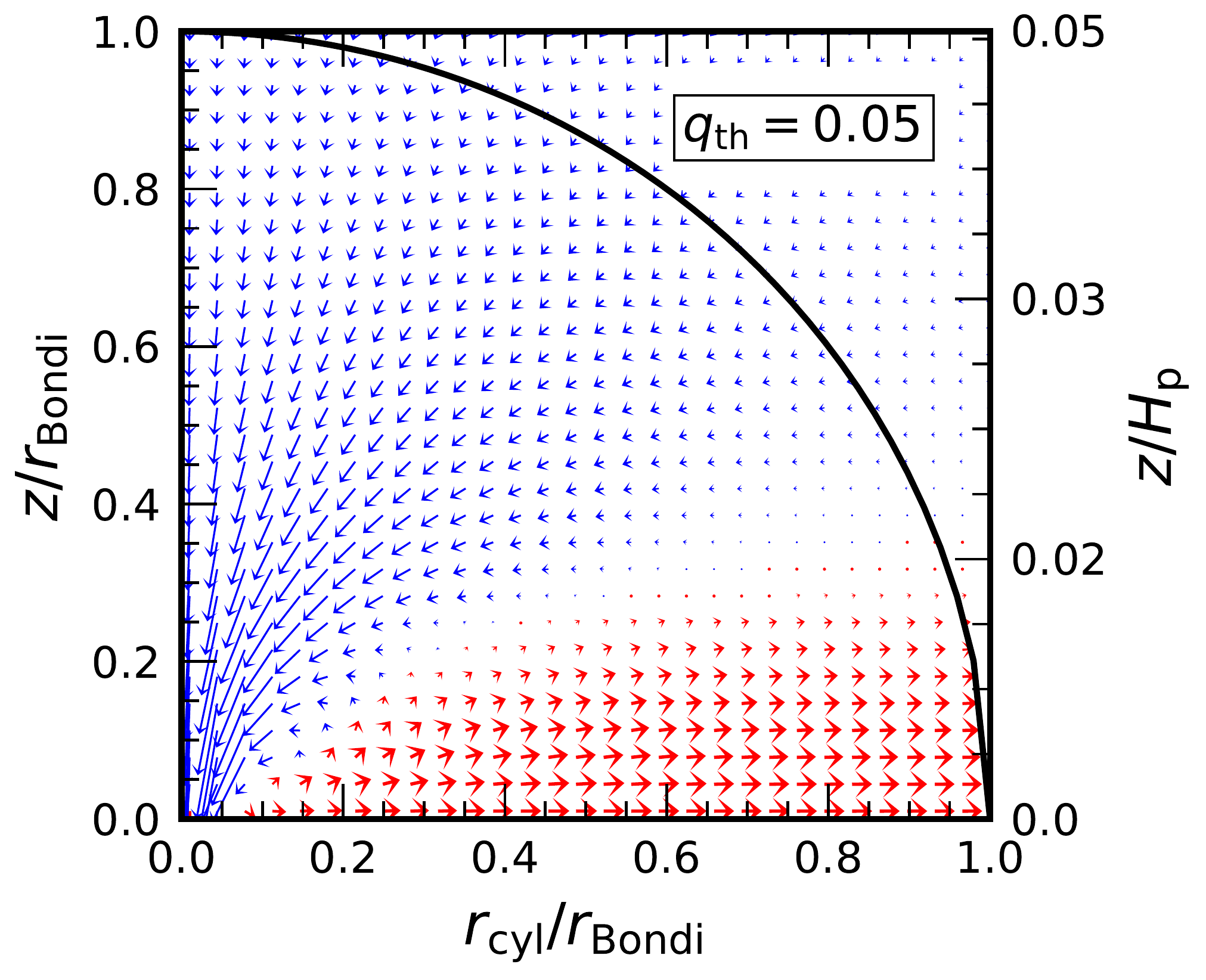}
\caption{\rev{Flow around a subthermal planet, located at $(r_{\rm cyl},\,z) = (0,0)$, with $\qth = 0.05$ and $\Hp/\Rp = 0.035$, from an $\athena$ simulation without using sink cells.} Data are time-averaged from $t = (10-15)\,\Omegap^{-1}$. Inflows 
(planet-centred radial velocity $v_r < 0$) 
are tagged blue and outflows are tagged red. The length of each arrow scales as the meridional gas velocity $\sqrt{v_z^2+v_{r_{\rm cyl}}^2}$, averaged over azimuth $\phi$, with the longest arrow 
having a magnitude of $6.5\cs$. The 
black curve marks the Bondi radius $r = \rb$.
\rev{Gas flows in along the planet's poles and, because the simulation does not include sink cells, exits through the midplane.}   }
  \label{fig:inflow_outflow}
\end{figure}

\begin{figure} 
\includegraphics[width=\columnwidth]{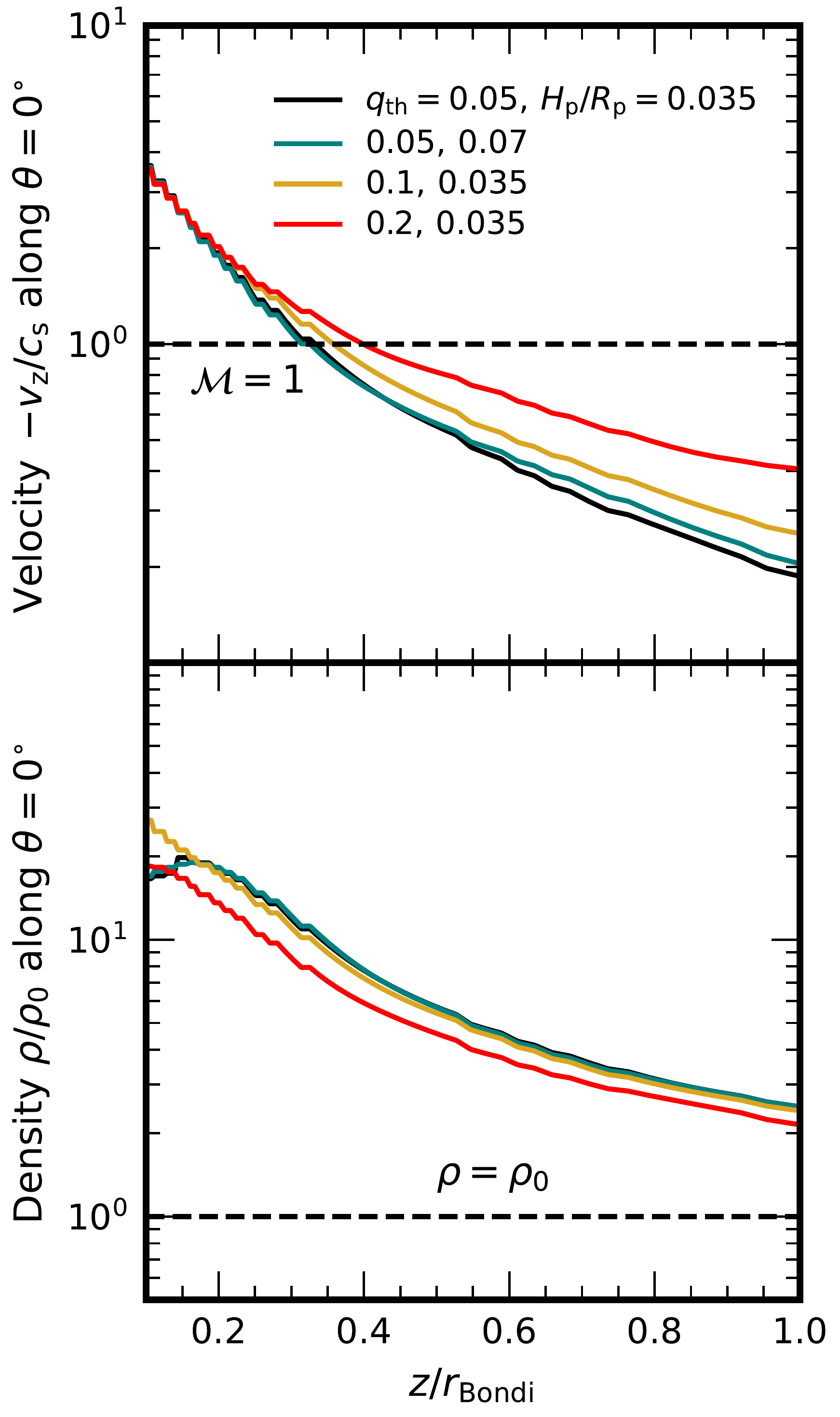}
\caption{ \rev{Time-averaged inflow velocity $-v_{\rm z}$ and density $\rho$ along the planet-centred $\theta = 0$ polar streamline, for $\qth \leq 0.2$, as measured with sink-less $\athena$ simulations. In all cases, the inflow becomes supersonic at $z \simeq 0.35\rb$, at which point $\rho \simeq 8\rho_0$.}   }
  \label{fig:pole}
\end{figure}

\rev{Figure \ref{fig:inflow_outflow} shows the meridional velocity field (in the $r_{\rm cyl}-z$ plane) around a subthermal planet in a simulation without any sink cells. Velocities have been averaged over azimuth $\phi$, and time-averaged from $t = 10\Omegap^{-1}$ to $15 \Omegap^{-1}$. In agreement with other studies that do not use sink cells \citep[][]{tanigawa_etal_2012, fung_etal_2015, szulagyi_etal_2016, bethune_rafikov_2019}, gas flows in along the planet's poles, from $\theta \simeq 60^\circ$ to $\theta = 0$ (blue arrows with $v_r < 0$).} Figure \ref{fig:pole} shows velocity and density along $\theta = 0$ for a few subthermal models. For $\qth = 0.05-0.2$, and independently of $\Hp/\Rp$, infalling gas achieves Mach 1 at $z \simeq 0.35\rb$ (Fig.~\ref{fig:pole}a), at which point $\rho \simeq 8\rho_0$ (Fig. \ref{fig:pole}b). \rev{Since these simulations do not include sink cells, gas eventually exits through the midplane (red arrows in Fig. \ref{fig:inflow_outflow}).}

\begin{figure} 
\includegraphics[width=\columnwidth]{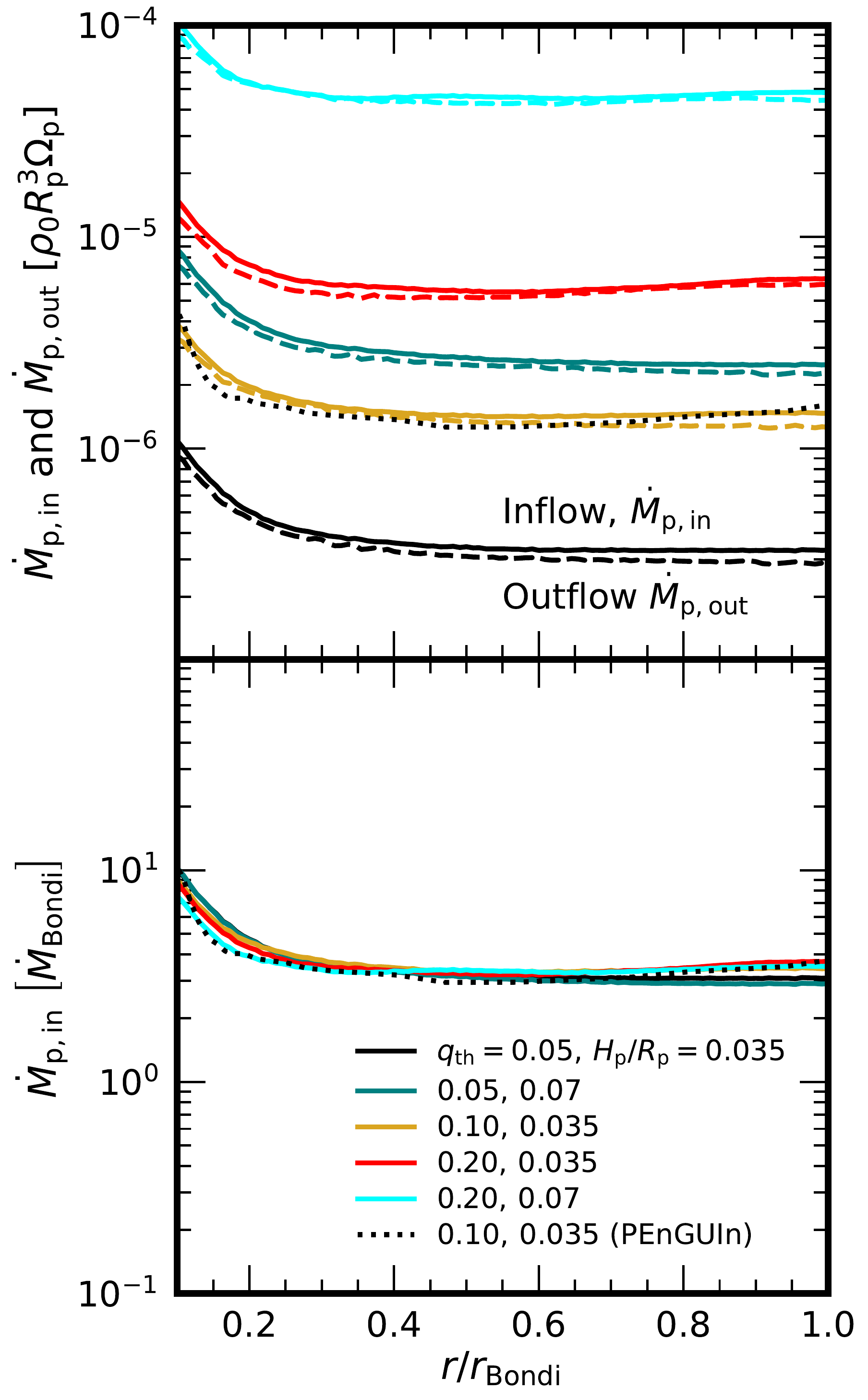}
\caption{ \rev{
\textit{Top:} Time-averaged mass inflow rates $\mdotin$ (solid lines) across planet-centred spheres of radius $r$ for subthermal planets, using simulations without sink cells.} Coloured lines show $\athena$ results for different input parameters, time-averaged from $t = (10-15) \Omegap^{-1}$. The dotted line is the inflow rate for a $\pgn$ simulation with $\qth = 0.1$ and $\Hp/\Rp = 0.035$, time-averaged from $t = (20-21) \times 2\pi\Omegap^{-1}$. 
We focus on the most steady region at $r \gtrsim 0.2\rb$ where each simulation converges to a value of $\mdotin$ that is nearly constant with $r$, and interpret this inflow rate as an upper limit to the planet's accretion rate. \rev{Since these simulations do not include sink cells to permanently accrete gas, outflow rates (dashed lines) balance inflow rates.}
\textit{Bottom:} Same as top, but showing only the inflow rates $\mdotin$ normalised by the Bondi rate $\mdotb = \rb^2 \rho_0 \cs$. }
\label{fig:Mdotr_subthermal_twopan}
\end{figure}

The top panel of Figure \ref{fig:Mdotr_subthermal_twopan} plots the time-averaged inflow rates $\mdotin(r)$ and outflow rates $\mdotout(r)$ (solid and dashed lines, respectively)
from the Bondi radius to inside of the sonic point for runs with various $\qth$ and $\Hp/\Rp$. Regions at $r \gtrsim 0.2 \rb$ are in a near-steady state, with inflow and outflow rates matching to within 15\%, and both nearly constant with $r$. At $r \lesssim 0.2\rb$, flow rates rise with decreasing $r$, implying by continuity that the density field here changes with time --- a consequence of the slight mismatch between inflow and outflow rates. Since this mismatch is less physical than numerical,  we focus on the more steady region at $r \gtrsim 0.2 \rb$ which offers a well-defined $\mdotin$ for every simulation. This inflow rate increases with $\qth$ and $\Hp/\Rp$, spanning two orders of magnitude across our parameter space. The bottom panel of Fig.~\ref{fig:Mdotr_subthermal_twopan} plots the same data in units of 
\begin{align}
\mdotb &\equiv \rb^2 \rho_0 \cs   \nonumber  \\ 
&= \qth^2 \left(\frac{\Hp}{\Rp}\right)^3 \rho_0 \Rp^3  \Omegap \,.
\label{eqn:mdotb}
\end{align}
\rev{So normalised, the time-averaged inflow rates for $\qth \leq 0.2$ and $0.2 < r/\rb < 1$ in sink-less $\athena$ and $\pgn$ runs collapse to
\begin{align}
\mdotin \simeq 3.5\mdotb \,.
\label{eqn:mdotb2}
\end{align}
}

\subsection{Superthermal limit}
\label{subsec:superthermal}

\begin{figure} 
\includegraphics[width=\columnwidth]{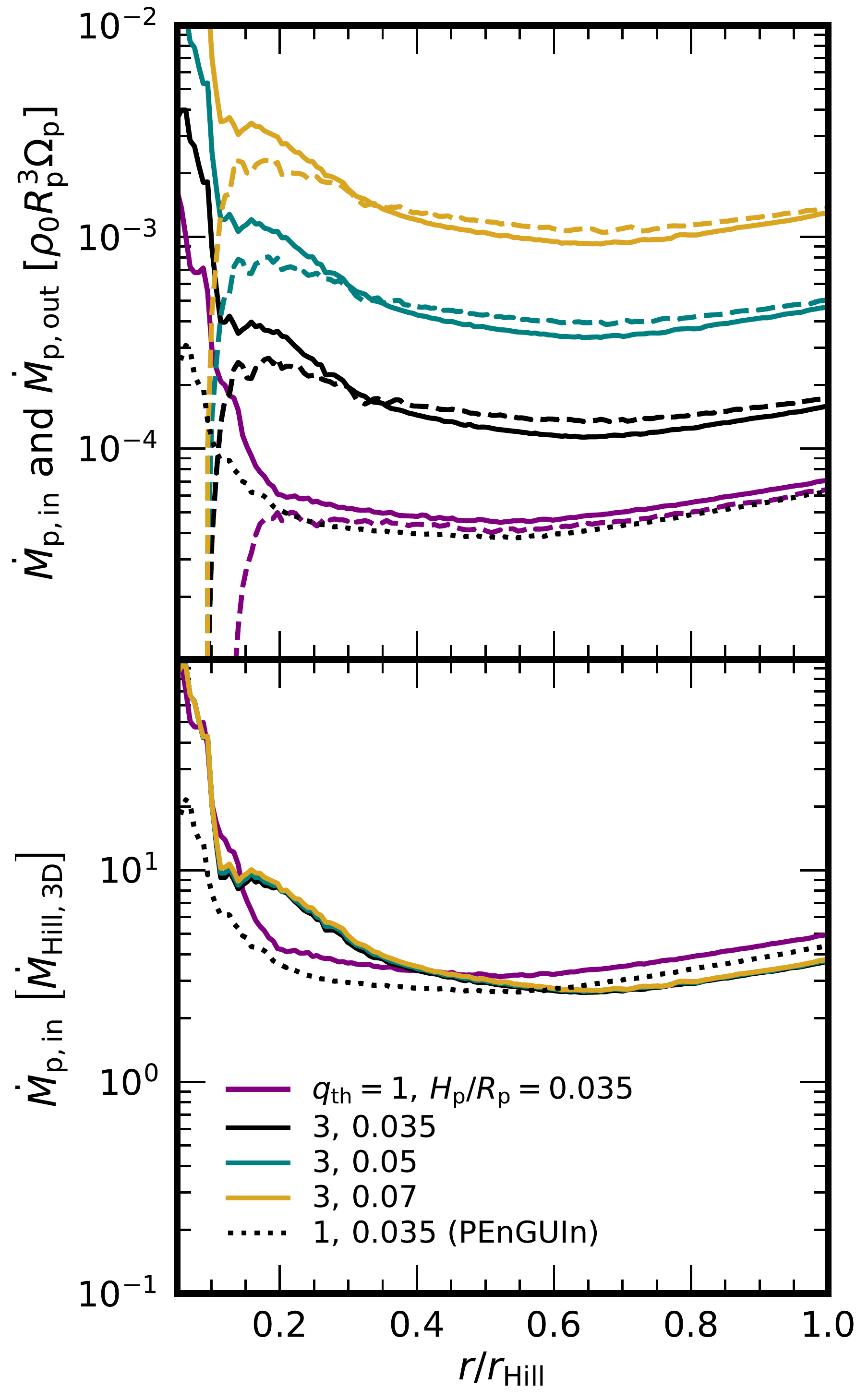}
\caption{ 
\textit{Top:} \rev{Mass inflow rates $\mdotin$ (solid lines) across planet-centred spheres of radius $r$ for marginally superthermal planets with $1 \leq \qth \leq 3$, using simulations without sink cells.} Coloured lines show $\athena$ results for different input parameters, time-averaged from $t=(10-15) \Omegap^{-1}$. The dotted line is the inflow rate for a $\pgn$ simulation with $\qth=1$ and $\Hp/\Rp=0.035$, time-averaged from $t=(20-21) \times 2\pi\Omegap^{-1}$. Just as subthermal runs have a nearly constant $\mdotin$ for $0.2 \lesssim r/\rb \lesssim 1$ (Fig.~\ref{fig:Mdotr_subthermal_twopan}), superthermal runs have a nearly constant $\mdotin$ between $0.4 \lesssim r/\rh \lesssim 1$ that we interpret as an upper limit to the planet's accretion rate.
\rev{Since these simulations do not use sink cells to permanently accrete gas, outflow rates (dashed lines) balance inflow rates. }
\textit{Bottom:} Same as top, but now showing only the inflow rates $\mdotin$ normalised by $\mdotthree = \rh^2 \times \Omegap \rh \times \rho_0$ (equation \ref{eqn:Mdot_3DH}).  
}
\label{fig:Mdotr_supthermal1}
\end{figure}

\begin{figure} 
\includegraphics[width=\columnwidth]{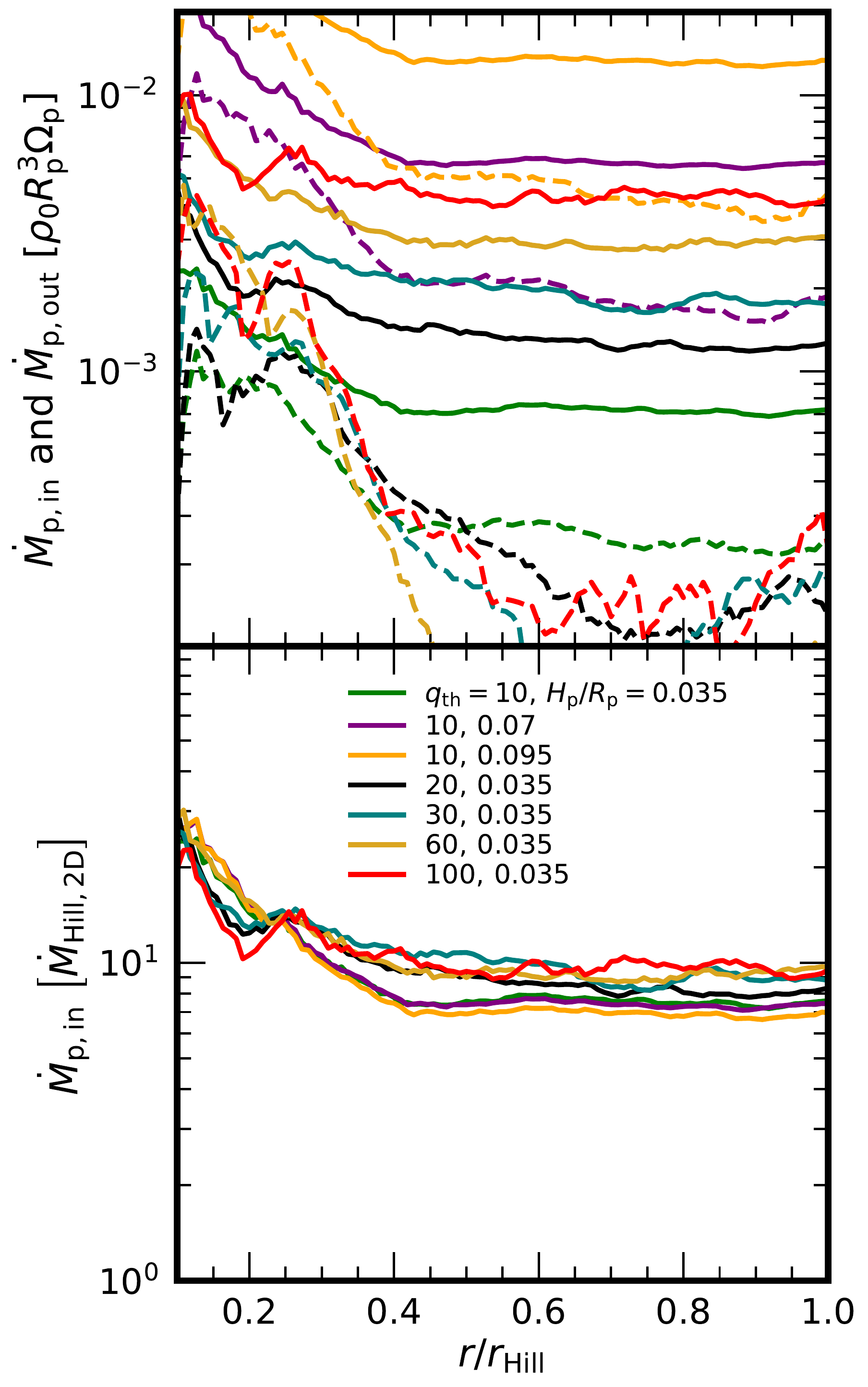}
\caption{ \textit{Top:} Same as the top panel of Figure \ref{fig:Mdotr_supthermal1} 
but for $\qth \geq 10$. Inflow rates remain nearly constant between $0.4 \lesssim r/\rh \lesssim 1$. At the times these data were taken (between 10 and $15\Omegap^{-1}$), $\mdotin$ has equilibrated but $\mdotout$ has not. We have checked in one case ($\qth = 10, \Hp/\Rp = 0.095$) that over longer runtimes $\mdotout$ grows to balance $\mdotin$.
\textit{Bottom:} Same as top, but now showing only the inflow rates $\mdotin$, normalised by $\mdottwo =  \rh \Hp \times \Omegap \rh \times \rho_0$ (equation \ref{eqn:Mdot_2DH}). 
}
\label{fig:Mdotr_supthermal2}
\end{figure}

As $\qth$ increases above 1, $\rb$ becomes larger than the planet's Hill radius:
\begin{align}
\rh &= \left(\frac{q}{3}\right)^{1/3}\Rp \nonumber \\ 
&= \left(\frac{1}{3}\right)^{1/3} \qth^{1/3} \Hp  \nonumber  \\
&= \left(\frac{1}{3}\right)^{1/3} \qth^{-2/3} \rb \,.
\end{align}
When $\rh < \rb$, stellar tidal forces are more important than thermal pressure in limiting how much gas can be gravitationally bound to the planet. Figures \ref{fig:Mdotr_supthermal1} and \ref{fig:Mdotr_supthermal2} show that for $\qth \geq 1$ there is a well-defined $\mdotin$ for $0.4 \lesssim r/\rh \lesssim 1$, motivating a Hill scaling for $\mdotin$ for superthermal planets by analogy with our earlier Bondi scaling for subthermal planets.
We start at $1 \leq \qth \leq 3$, in the ``3D'' regime where the Hill sphere is still embedded in the circumstellar disc ($\rh < \Hp$). Here the Hill sphere presents a cross-sectional area of $\sim$$\rh^2$ to gas shearing toward it at speed $\sim$$\Omegap \rh$. The inflow rate then scales as
\begin{align}
\mdotthree &\equiv \rh^2 \times  \Omegap \rh \times \rho_0 \nonumber \\ 
 &= \frac{\qth}{3}\left(\frac{\Hp}{\Rp}\right)^3 \rho_0 \Rp^3 \Omegap \,,
 \label{eqn:Mdot_3DH}
\end{align}
a weaker dependence on planet mass than $\mdotb \propto \qth^2$. \rev{The bottom panel of Fig.~\ref{fig:Mdotr_supthermal1} confirms the expected scaling, showing that for $0.4 \lesssim r/\rh \lesssim 1$ and $1 \leq \qth \leq 3$, our data from sink-less $\athena$ and $\pgn$ simulations collapse to}
\begin{equation}
\mdotin \simeq 4\mdotthree \,. 
\label{eqn:Mdot_3DH_coeff}
\end{equation}

\begin{figure} 
\hspace{-0.2cm}\includegraphics[width=1.1\columnwidth]{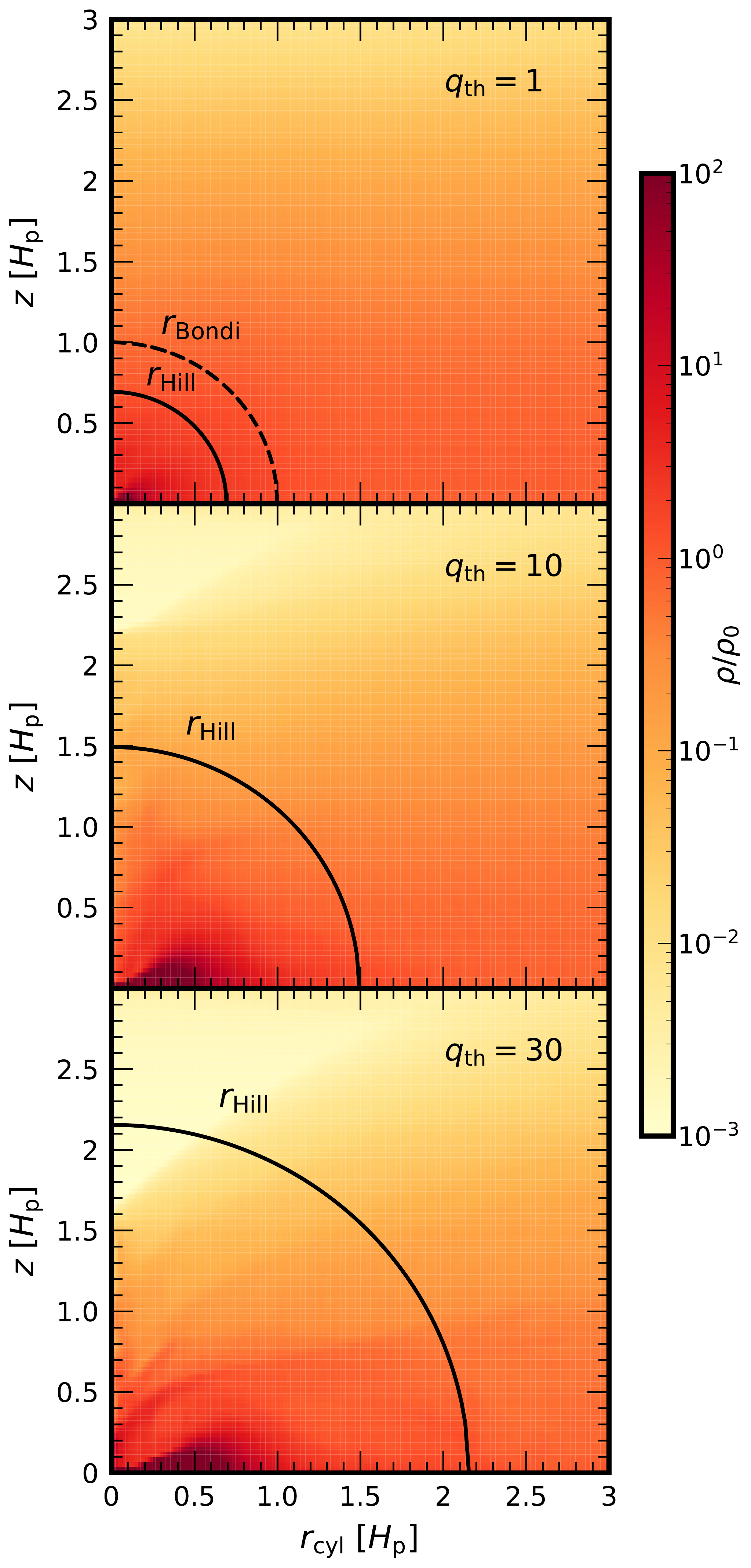}
\vspace{-0.2in}
\caption{Meridional slices of the density field around superthermal planets, taken at $t = 10 \Omegap^{-1}$ and azimuthally averaged, from \rev{sink-less} $\athena$ runs with $\Hp/\Rp = 0.035$ and $\qth$ increasing from top to bottom. The planet is at the origin $(r_{\rm cyl},\,z)=(0,0)$ of each panel. In the superthermal regime we measure inflow rates through the Hill sphere (solid arc) because it is smaller than the Bondi sphere (dashed arc; for $\qth = 10$ and 30, the Bondi sphere expands out of the plotted range).  For $\qth = 1$, the Hill sphere is still immersed in the disc ($\rh < \Hp$); inflowing gas covers the Hill sphere's entire cross section $\sim$$\rh^2$, with an accretion rate given by equation \ref{eqn:Mdot_3DH}.
For $\qth = 30$, the Hill sphere ``pops out'' of the disc ($\rh > \Hp$). Now the gas that enters the Hill sphere from above the planet is much less dense than the gas that enters through the midplane; the cross section  presented by gas to the Hill sphere is $\sim$$\rh \Hp$, with an accretion rate given by equation \ref{eqn:Mdot_2DH}.
}
  \label{fig:xz}
\end{figure}

When $\qth \gtrsim 10$, the Hill sphere ``pops out'' of the circumstellar disc ($\rh > \Hp$), as illustrated in Figure \ref{fig:xz}. The density near the Hill sphere's pole is so low that the inflow comes mostly from the midplane; accretion is now more 2D. Midplane gas presents a cross-sectional area to the Hill sphere of $\sim$$\rh \Hp$ and flows in at a rate
\begin{align}
\mdottwo &\equiv \rh \Hp \times  \Omegap \rh \times \rho_0 \nonumber \\ 
 &= \left(\frac{\qth}{3}\right)^{2/3}\left(\frac{\Hp}{\Rp}\right)^3 \rho_0 \Rp^3 \Omegap \,,
\label{eqn:Mdot_2DH}
\end{align}
which scales even more weakly with planet mass than $\mdotthree$.
\rev{The bottom panel of Fig.~\ref{fig:Mdotr_supthermal2} shows that for $0.4 \lesssim r/\rh \lesssim 1$ and $\qth \geq 10$, our data from sink-less $\athena$ 
simulations collapse to
}
\begin{align}
\mdotin \simeq 9\mdottwo \,.
\label{eqn:Mdot_2DH_coeff}
\end{align}
We find that for larger $\qth$ the outflow rate $\mdotout$ equilibrates more slowly than $\mdotin$.
The data for Fig. \ref{fig:Mdotr_supthermal2} were taken when $\mdotin$ had equilibrated but $\mdotout$ had not. \rev{We have checked for $\qth = 10$ and $\Hp/\Rp = 0.095$ that when the simulation is extended to 100$\Omegap^{-1}$, outflow grows to match inflow, as expected for sink-less runs.}

\begin{figure} 
\includegraphics[width=\columnwidth]{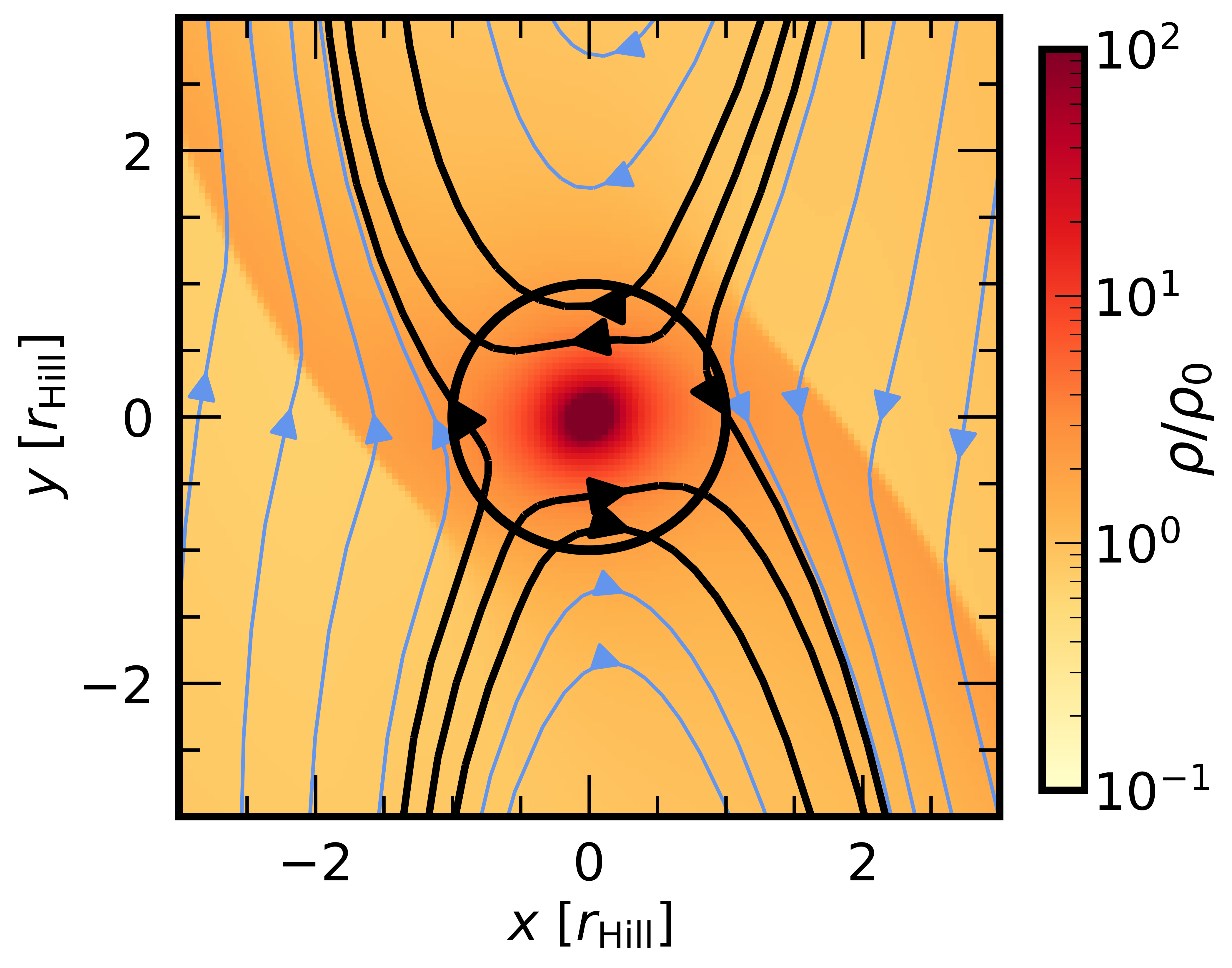}
\caption{Gas streamlines and density in the disc midplane at $t = 10\,\Omegap^{-1}$ from a \rev{sink-less} $\athena$ simulation with $\qth = 1$ and $\Hp/\Rp = 0.035$. Data are in Cartesian coordinates centred on the planet, where $x$ points away from the star and $y$ points along the planet's orbit.
\rev{The Hill sphere (black circle) has gas fed into it by streamlines colored black; many of these streamlines are on horseshoe orbits (top and bottom), while others are circulating (sides). The rate at which these streamlines carry mass into the sphere defines $\mdotin (\rh)$, and the rate at which they carry mass out defines $\mdotout (\rh)$. Since the simulation shown here does not include sink cells, $\mdotout \approx \mdotin$. In analogous simulations that do use sink cells (section \ref{subsec:sink}), $\mdotout \ll \mdotin$, while $\mdotin$ remains within a factor of 3 of its value derived without sink cells. For this figure we highlight $\rh$ as the boundary across which we measure mass fluxes; in Figs.~\ref{fig:Mdotr_subthermal_twopan}, \ref{fig:Mdotr_supthermal1}, \ref{fig:Mdotr_supthermal2}, and \ref{fig:sink}, we vary the measurement boundary by a factor of 10, and also consider $\rb$ as an alternative reference boundary.
}}

  \label{fig:xy}
\end{figure}

Figure \ref{fig:xy} plots gas streamlines in the disc midplane around a $\qth = 1$ planet. Most of the material that crosses the Hill sphere is sourced by a subset of horseshoe orbits flowing in from either side of the planet's orbit (see also fig. 4 of \citealt{lubow_etal_1999}; fig. 3 of \citealt{tanigawa_watanabe_2002}). \rev{Since the simulation does not include sink cells, nearly all of the inflowing gas also exits the Hill sphere, so that $\mdotout \approx \mdotin$. 
}

\subsection{Gaps}
\label{subsec:gaps}
\begin{figure*}
\includegraphics[width=1.05\textwidth]{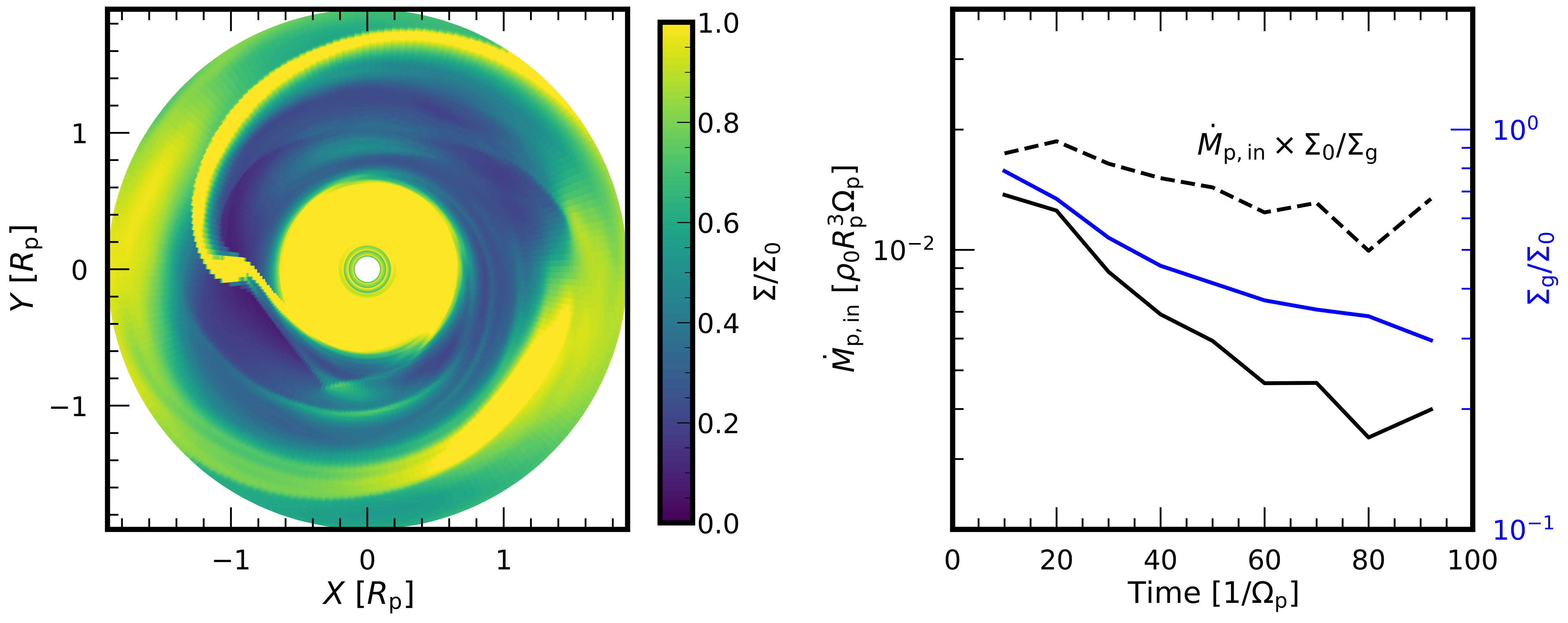}

\caption{Effects of gap opening on inflow rates, demonstrated using our \rev{sink-less} simulation with $\qth = 10$ and $\Hp/\Rp = 0.095$. 
The left panel shows a snapshot of the gas surface density $\Sigma$ in the disc midplane at $t = 100\,\Omegap^{-1}$. Data are in Cartesian coordinates centred on the star and the planet is at $(X,Y) = (-1,0)$. The colour scale is capped at $\Sigma_0$, the initial surface density at the planet's position. We compute a spatially averaged surface density $\Sigg$ between $\Rp - \rh < R <  \Rp + \rh$, excluding the circumplanetary region $\Psi_{\rm p} - 2\rh/\Rp < \Psi < \Psi_{\rm p} + 2\rh/\Rp$. The right panel shows that $\Sigg$ decreases as the simulation progresses (solid blue curve read using the right-hand axis) and that the inflow rate through the Hill sphere $\mdotin$ (solid black curve, left-hand axis) tracks this decline, as expected because the planet is fed by material in the gap (and not from the overdense spirals seen in the left panel; see also Fig.~\ref{fig:xy}). The inflow rate re-normalised by $\Sigma_0/\Sigg$ is more constant with time (dashed black curve, left-hand axis).
}
  \label{fig:gaps}
\end{figure*}

The inflow rates in Figs. \ref{fig:Mdotr_subthermal_twopan}-\ref{fig:Mdotr_supthermal2} were time-averaged between $t = (10-15) \,\Omegap^{-1}$, before the planets have cleared gaps around themselves. Since the planet is fed by co-orbital material (Fig. \ref{fig:xy}), we expect that inflow rates should scale in proportion to the surface density in the gap, a.k.a.~the gap depth. To test this, we extend the runtime of our $\qth = 10$, $\Hp/\Rp = 0.095$ simulation to 100$\Omegap^{-1}$ which allows gaps to develop more fully. The left panel of Figure \ref{fig:gaps} shows the gap carved by the planet at the end of this extended simulation.

We compute the average surface density in the gap $\Sigg$ by summing the mass in all cells in an annulus with $\Rp - \rh < R <  \Rp + \rh$, excluding those in the circumplanetary region with $\Psi_{\rm p} - 2\rh/\Rp < \Psi < \Psi_{\rm p} + 2\rh/\Rp$, and dividing by the surface area of the excised annulus. The right panel of Fig. \ref{fig:gaps} shows that the decline of $\Sigg$ over the simulation duration (solid blue curve) is roughly paralleled by the decline in $\mdotin$ through the Hill sphere (solid black curve), and \rev{that $\mdotin$ re-normalised by the gap depth can describe the actual inflow rate to within a factor of 2 (dashed black curve). This result also agrees with \citet[][their fig. 21]{fung_etal_2019} who showed that the average surface density in the circumplanetary region (i.e., the region we excised to compute $\Sigg$) scales in proportion to $\Sigg$.}

\rev{Thus we expect that equations \ref{eqn:mdotb2}, \ref{eqn:Mdot_3DH_coeff}, and \ref{eqn:Mdot_2DH_coeff} for planet inflow rates can still be used in the presence of gaps, with $\rho_0$ in those equations set equal to the midplane density averaged over the annular gap, excluding the region nearest the planet.\footnote{\rev{This procedure sidesteps having to specify disc viscosity as it is encoded in the gap depth \citep[e.g.][]{duffell_macfadyen_2013, fung_etal_2014, kanagawa_etal_2015}. Our simulations do not include an explicit viscosity. Including one would presumably lead to accretion of circumplanetary material onto the planet, reducing $\mdotout$ but leaving $\mdotin$ unchanged.}}
}

\subsection{Sink cell runs}
\label{subsec:sink}
\begin{figure}
\includegraphics[width=\columnwidth]{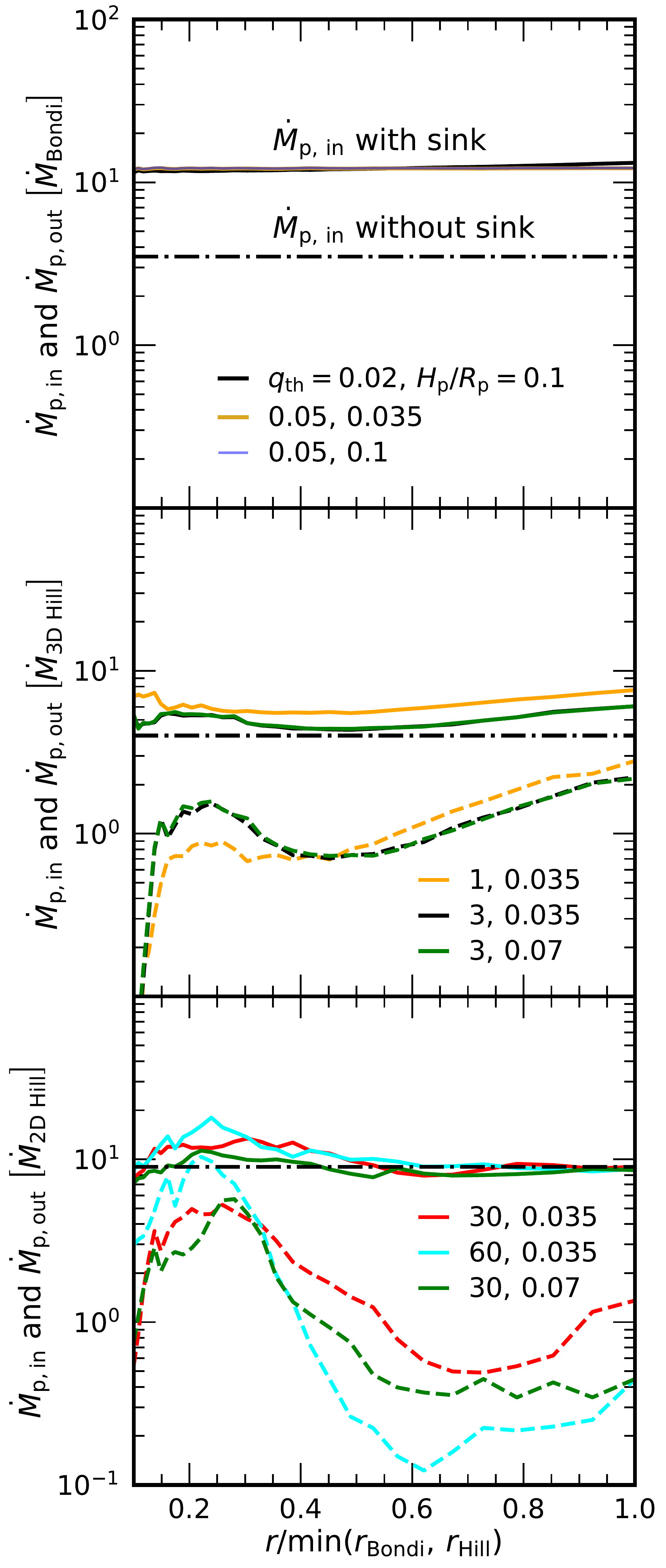}
\vspace{-0.55cm}
\caption{ \rev{Mass inflow rates $\mdotin$ (solid curves),  
averaged over $t = 10-15 \Omegap^{-1}$, from $\athena$ simulations using sink cells, in the subthermal (top panel), marginally superthermal (middle panel), and superthermal (bottom panel) regimes, plotted in the scaled units appropriate to those limits (equations \ref{eqn:mdotb}, \ref{eqn:Mdot_3DH}, \ref{eqn:Mdot_2DH}, respectively). These scaled units and their dependencies on $\qth$ and $\Hp/\Rp$ were motivated by runs without sink cells, but are seen here to apply just as well to runs with sink cells, aside from order-unity differences in normalisation; compare solid curves to the dash-dot black lines summarising our sink-less results (equations \ref{eqn:mdotb2}, \ref{eqn:Mdot_3DH_coeff}, \ref{eqn:Mdot_2DH_coeff}) to see that inflow rates with sink cells are up to $3\times$ higher than rates without sinks. Outflow rates with sinks (dashed curves) are markedly lower than inflow rates; the subthermal runs have $\mdotout = 0$.}
}
  \label{fig:sink}
\end{figure}

\begin{figure} 
\includegraphics[width=\columnwidth]{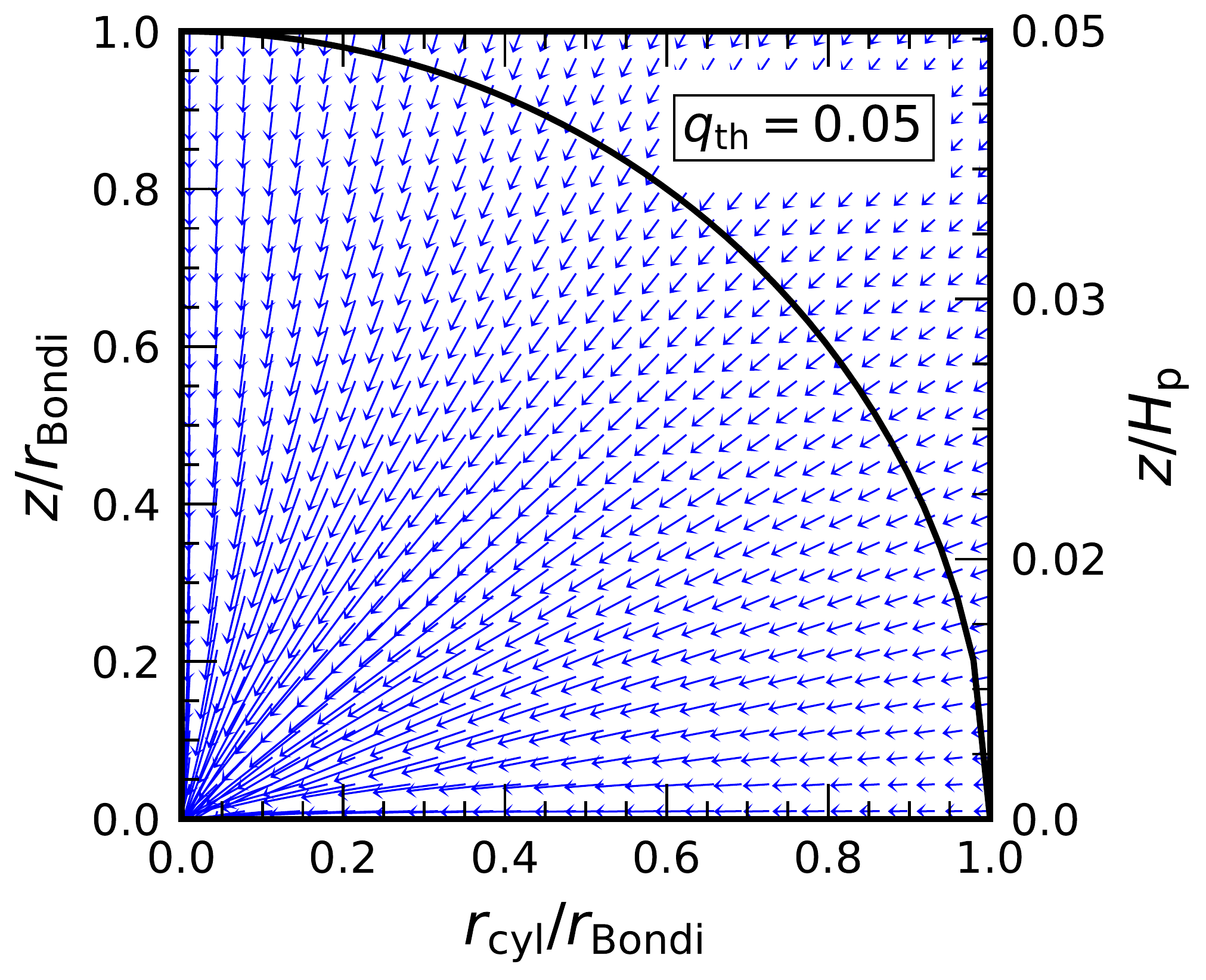}
\caption{\rev{Time-averaged flow 
($t = 10-15 \Omegap^{-1}$) 
around a subthermal planet at $(r_{\rm cyl},\,z) = (0,0)$, with $\qth = 0.05$ and $\Hp/\Rp = 0.035$, from an $\athena$ simulation including sink cells. The length of each arrow scales as the meridional gas velocity $\sqrt{v_z^2+v_{r_{\rm cyl}}^2}$, averaged over azimuth $\phi$, with the longest arrow having a magnitude of $6.2\cs$. The black curve marks $r = \rb$. Using sink cells eliminates the midplane outflow found in sink-less subthermal simulations (Fig.~\ref{fig:inflow_outflow}). Gas accretion is here nearly spherically symmetric, which boosts inflow rates $\mdotin$ compared to the sink-less case.
}
}
  \label{fig:inflow_outflow_sink}
\end{figure}

\rev{Figure \ref{fig:sink} plots $\mdotin$ vs. $r$ from $\athena$ simulations that use sink cells near the planet. Like their sink-less counterparts, these runs show a 
well-defined 
$\mdotin$ for $0.1\,\lesssim r/\minrh \lesssim 1$ across all three subthermal, marginally superthermal, and superthermal regimes. Fig.~\ref{fig:sink} also shows that $\mdotin$ simulated with sink cells follows the same scalings with $\qth$ and $\Hp/\Rp$ that we identified from runs without sink cells (equations \ref{eqn:mdotb}, \ref{eqn:Mdot_3DH}, \ref{eqn:Mdot_2DH}). Overall magnitudes for $\mdotin$ are also similar, with the largest difference 
in the subthermal limit where $\mdotin$ is $3\times$ higher with sink cells than without. This higher inflow rate is within 15\% of the classic Bondi accretion rate onto a point mass from spherically symmetric, isothermal gas: $\mdot = 4.48\pi G^2\Mp^2 \rho_0/\cs^3$ (table 1 of \citealt{bondi_1952}). The flow field around a subthermal planetary sink (Figure \ref{fig:inflow_outflow_sink}) is nearly spherically symmetric and lacks the midplane outflow of non-sink simulations (Fig.~\ref{fig:inflow_outflow}). }

\vspace{3cm}

\section{Summary and Discussion}
\label{sec:summary_discussion}

\begin{figure*} 
\includegraphics[width=\textwidth]{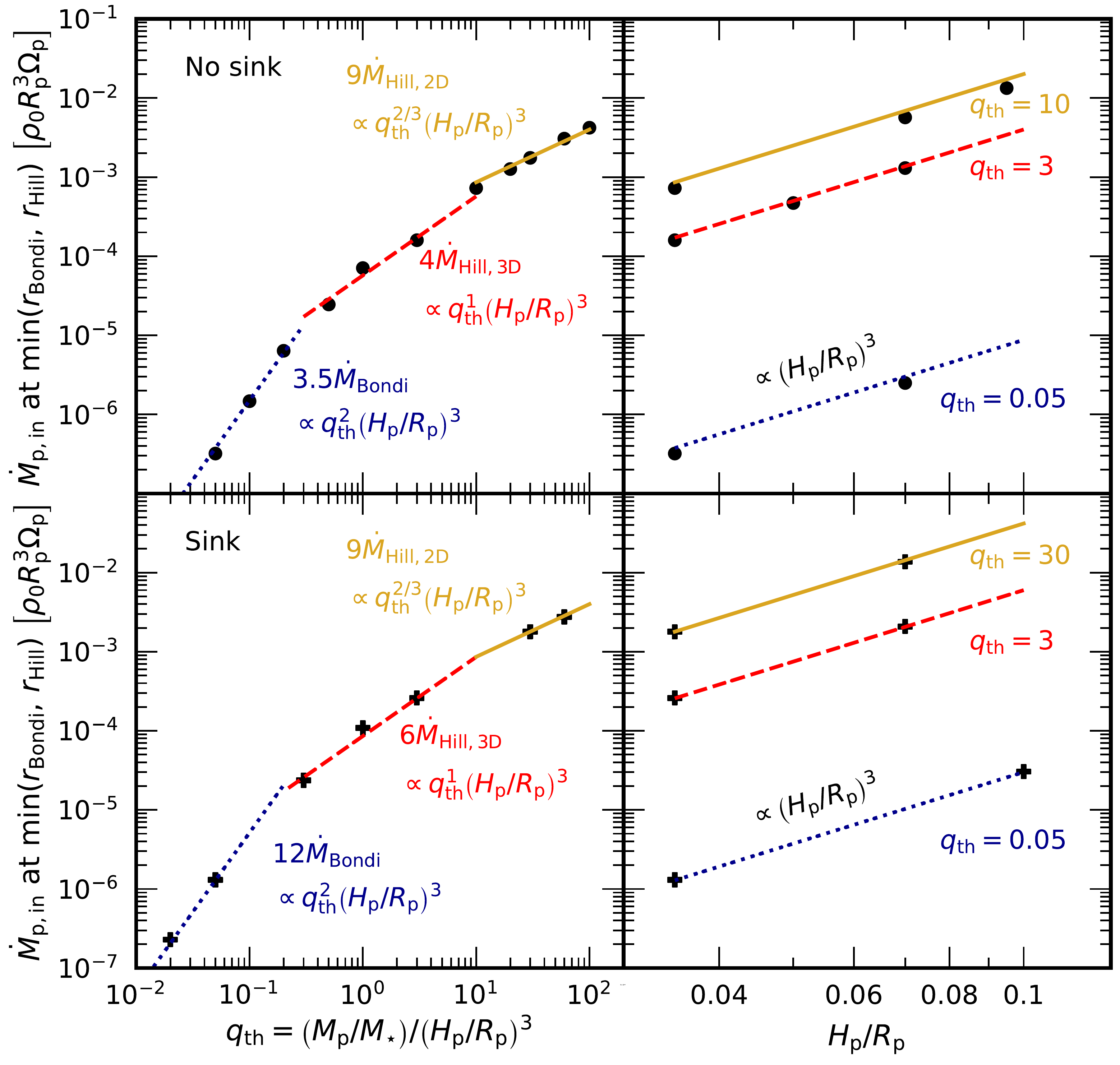}
\caption{
\rev{ How a planet's maximum accretion rate $\mdotin$ scales with planet mass and disc properties.  \textit{Top row:} The left panel varies the thermal mass parameter $\qth  = \left(\Mp/\Mstar\right)/\left(\Hp/\Rp \right)^{3}$ at fixed $\Hp/\Rp = 0.035$, and the right panel varies $\Hp/\Rp$ for three values of $\qth$. Filled circles and open squares show results from sink-less $\athena$ and $\pgn$ simulations, respectively. Coloured curves are power-law scalings from order-of-magnitude arguments, adjusted in normalisation to match the simulation data (equations \ref{eqn:sum1}-\ref{eqn:sum3}). Inflow rates follow the $\qth^2$ scaling for Bondi accretion when $\qth \lesssim 1$ (dotted blue curves). For larger $\qth$, stellar tides restrict the planet's reach to $\rh < \rb$, and $\mdotin$ scales less strongly with $\qth$. Such Hill accretion is 3D (dashed red curves) when $\rh < \Hp$, and 2D for the largest values of $\qth$ when $\rh > \Hp$ (solid gold curves). \nick{For planets orbiting within gaps, all of these rates should be scaled down in proportion to the gap depth.} 
\textit{Bottom row:} Same as top but for simulations that use sink cells near the planet. Sink-cell inflow rates follow the same analytic scalings with $\qth$ and $\Hp/\Rp$ as sink-less rates, but have normalisations up to 3$\times$ higher. }
} 
  \label{fig:Mdot_master}
\end{figure*}
Using global, isothermal, 3D hydrodynamic simulations, we measured the maximum accretion rate of a planet embedded in a gaseous circumstellar disc. This upper bound is given by $\mdotin$, the rate at which gas enters the planet's gravitational sphere of influence, which is the smaller of the planet's Bondi and Hill spheres. We would like to know how much of the inflowing gas becomes permanently bound, but this cannot be determined without knowing how the gas sheds angular momentum, or stays cool against adiabatic compression or shock heating; this physics is not captured in our inviscid, isothermal simulations. The upper limit we have established is relevant for protoplanets of at least several Earth masses with self-gravitating gas envelopes, 
accreting in the hydrodynamic runaway or post-runaway regimes  
(e.g.~\citealt{ginzburg_chiang_2019a, ginzburg_chiang_2019b}).

\rev{Figure \ref{fig:Mdot_master} summarises our results. The planet's
thermal mass parameter $\qth$ controls the geometry and magnitude of
inflow according to:
\begin{numcases}{ \frac{\mdotin}{\rhog \Omegap \Rp^3} \simeq }
  C_1\qth^2 \left(\frac{\Hp}{\Rp}\right)^3  
  \,\,\,\,\,\,\,\,\,\,\,\,\,\,\,\, \qth 
  \lesssim 0.3
  \,\,
 \label{eqn:sum1} \\
 \nonumber  \\
         C_2\qth \left(\frac{\Hp}{\Rp}\right)^3  \,\,\,\, \,\,\,\,\,\,\,\,\,\,\,\, 
         0.3 \lesssim 
         \qth 
         \lesssim 10 
         \label{eqn:sum2} \,\,
         \\ \nonumber \\ 
          C_3
          \qth^{2/3} \left(\frac{\Hp}{\Rp}\right)^3 \,\,\,\,\,\,\,\,\,\,\,\,\, \qth 
          \gtrsim 10 
          \label{eqn:sum3} \,\, 
\end{numcases}
where $\qth \equiv \left(\Mp/\Mstar \right)\left(\Hp/\Rp \right)^{-3}$, $\Mp$ and
$\Mstar$ are the planet and star masses, and $\rhog$, $\Hp/\Rp$, and
$\Omegap$ are the ambient midplane gas density, disc aspect ratio, and
Keplerian angular frequency at the planet's orbital radius $\Rp$. When we model the planet with sink cells, then the constants $\{C_1,\,C_2,\,C_3\} = \{12,2,9/3^{2/3}\}$; otherwise  $\{C_1,\,C_2,\,C_3\} = \{3.5,4/3,9/3^{2/3}\}$.} 
All of these constants, including the $\qth$ boundary values separating the three regimes, are calibrated from simulations.

For subthermal planets with $\qth \lesssim 0.3$, gas flows in at a Bondi-like rate, increasing as the square of the planet mass. Superthermal inflow rates scale more weakly with planet mass because stellar tides restrict the planet's reach for $\qth \gtrsim 0.3$, and because the Hill sphere \nick{pops well out of the disc for $\qth \gtrsim 10$.} 
Whereas the (minimum) mass doubling time $\Mp/\mdotin$ at fixed $\rhog$ decreases with planet mass in the strongly subthermal regime (i.e.~growth is potentially super-exponentially fast), the doubling time increases with planet mass in the strongly superthermal regime (power-law growth). This last result should help to limit the masses to which planets can grow (e.g.~\citealt{rosenthal_etal_2020}).

\rev{In equations \ref{eqn:sum1}-\ref{eqn:sum3}, $\rhog$ is the disc density outside the planet's immediate sphere of influence but still within the planet's horseshoe co-orbital region. This density is lowered as the planet opens a gap about its orbit. We have checked that the planet's inflow rate simply scales in proportion to the gap surface density, which follows its own scalings with $M_{\rm p}/M_\star$, $\Hp/\Rp$, and dimensionless viscosity $\alpha$ \citep[e.g.][]{duffell_macfadyen_2013, fung_etal_2014, kanagawa_etal_2015}. These gap scalings can be combined with the scalings we have established in this paper to determine how inflow rates scale in the net. For example, for subthermal planets that open deep gaps (which they can if $\alpha$ is small enough), $\rhog \propto \Mp^{-2}$, and therefore $\mdotin \propto \rhog \qth^2 \propto \Mp^0$.
}

\subsection{Comparison with other simulations}
\label{subsec:comparison}
\rev{For the most part our results confirm or can be reconciled with previous calculations.} 
We found that inflow rates scale with the smaller of the Bondi and Hill spheres. In their study of orbital migration, \cite{masset_etal_2006} determined that the smaller of the two regions also matters for the torque exerted by the disc, and that the width of the horseshoe zone changes its dependence on planet mass at $\qth \approx 0.5$ (see their fig.~9), similar to where we found a break in the inflow scaling.

\rev{In the subthermal $\qth \lesssim 0.3$ regime, the 3D, isothermal, sink-cell simulations of \cite{dangelo_etal_2003} and \cite{machida_etal_2010} (compiled in fig.~1 of \citealt{tanigawa_tanaka_2016}) appear consistent with a Bondi accretion rate scaling, $\mdotin \propto \Mp^2$, as we found. When our respective subthermal inflow rates are scaled to the same disc parameters ($\Hp/\Rp = 0.05$, $\Rp = 5.2$ au, and an unperturbed background disc density of $\rho_{\rm g} = 1.4 \times 10^{-11} \rm g/cm^3$), their rates are about an order of magnitude lower than what our equation \ref{eqn:sum1} predicts using $C_1 = 12$.}


\nick{\cite{bethune_rafikov_2019} studied planets with $0.5 \leq \qth \leq 4$ in the marginally superthermal regime using 3D sink-less, isothermal, and inviscid simulations. Their simulations do not use a softened potential and instead model the planet's core as an impermeable surface. They report some permanent accretion of gas because of dissipation in standing shocks near this core. Encouragingly, their net mass accretion rate $\mdot = \mdotin - \mdotout$ grows linearly with $\Mp$ and is independent of $\Hp/\Rp$, matching the scalings in our equation \ref{eqn:sum2} for $\mdotin$ (see their fig.~12 and equation 13; they do not give the breakdown of inflow vs. outflow). Their net rate is 15$\times$ lower than our sink-less inflow rate, possibly because only a narrow set of polar streamlines intersects the core and permanently accretes via shocks (see the cyan curve in their fig. 2 marking the width of the shocked region). As in our sink-less runs, most of the material entering their simulated Hill spheres exits through the midplane.
}

\cite{tanigawa_watanabe_2002} also considered the marginally superthermal regime. For $0.5 < \qth < 6$, they found a steeper $\Mp^{4/3}$ scaling
for the accretion rates onto their planetary sink cells.\footnote{\nick{
\cite{tanigawa_watanabe_2002} use the normalised sound speed $\tilde{C}_{\rm iso} = \Hp/\rh $ to describe their simulated planets. We translate their values using $\qth = 3/\tilde{C}_{\rm iso}^3$. }} But this result
is based on 2D (vertically integrated) simulations, in a regime where
accretion is actually more 3D (\citealt{bethune_rafikov_2019}, and our section \ref{subsec:superthermal}). 
We expect better agreement between 2D and 3D simulations when $\rh
\gtrsim \Hp$ ($\qth \gtrsim 10$). The self-gravitating gas clumps
modeled in 2D as sink cells by \cite{zhu_etal_2012} fall into this
fully superthermal limit, and have accretion rates which match equation \ref{eqn:sum3} in magnitude and scaling (see their equation 15).

\subsection{Connecting to observations}
\label{subsec:connect_to_obs}
We use our results for $\mdotin$ to place lower bounds on the growth
timescales for observed or suspected protoplanets embedded in
circumstellar gas discs. Table \ref{tab:data} updates the compilation
of \cite{choksi_chiang_2022} of such planets, listing their possible
masses $\Mp$ and, where optically thin C$^{18}$O data are available,
ambient gas surface densities $\Sigg$ (for details, see the caption to
Table \ref{tab:data}, Appendix \ref{sec:data}, and
\citealt{choksi_chiang_2022}). From $\Sigg$ we compute
$\rhog = \Sigg/\left(\sqrt{2\pi} H_{\rm p}\right)$ (assuming the disc
is isothermal and in hydrostatic equilibrium) and from there a
planet's minimum mass-doubling timescale
$\min \, (t_{\rm double} ) = \Mp/\mdotin$ (column 10 of Table
\ref{tab:data}) \rev{using equations \ref{eqn:sum1}-\ref{eqn:sum3} with the larger coefficients from our sink-cell simulations.}

\begin{figure*} 
\includegraphics[width=0.98\textwidth]{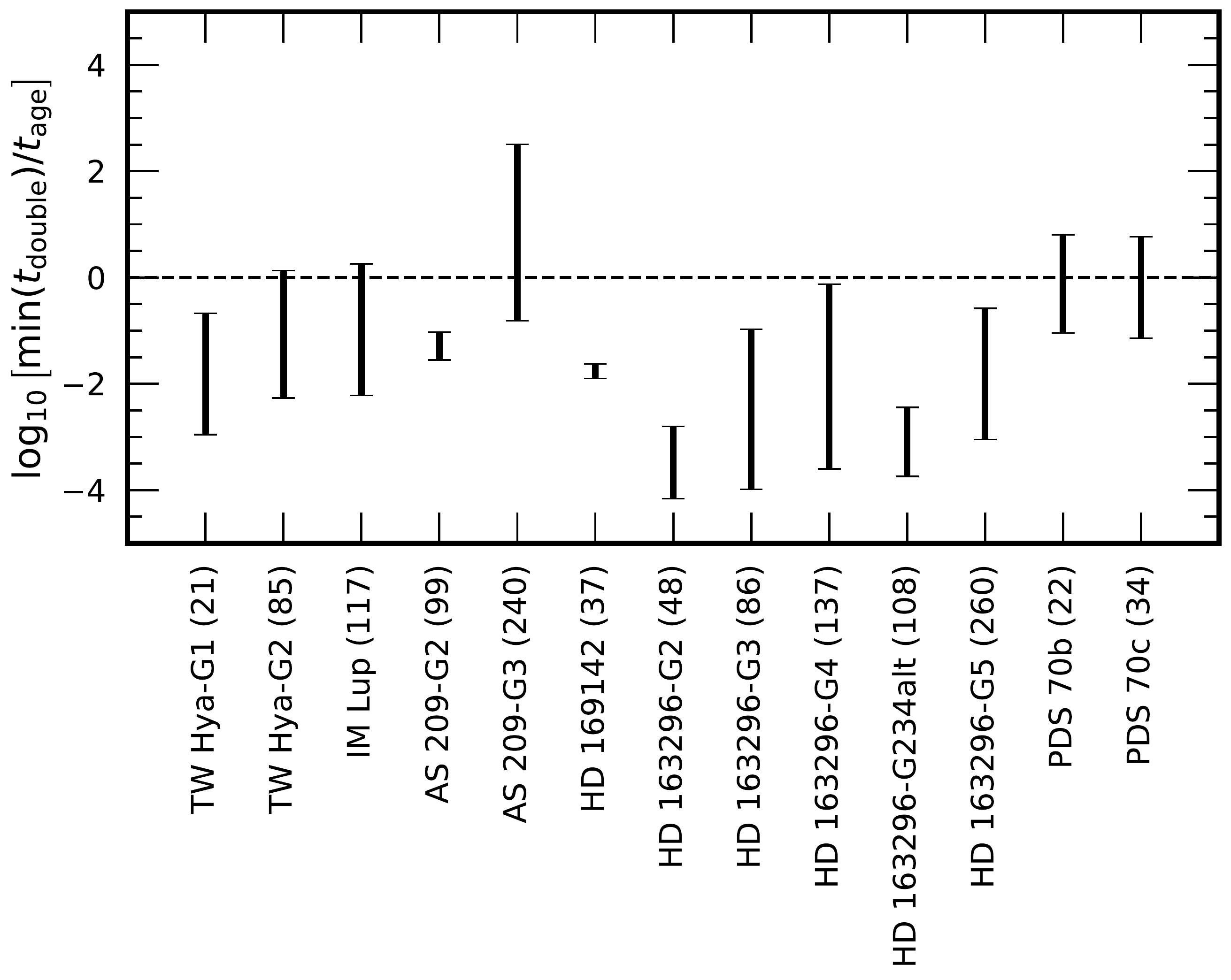}
\caption{The minimum mass-doubling time $\mtd = \Mp/\mdotin$, divided by the system age $t_{\rm age}$, for the subset of hypothesised planets in Table \ref{tab:data}  (confirmed in the case of PDS 70) having gap surface densities $\Sigg$ from C$^{18}$O emission. System names on the horizontal axis are followed by ``G\#'' to identify a specific gap in systems with multiple gaps, and by the planet's orbital radius in units of au in parentheses. The ``HD 163296-G234alt'' entry corresponds to a scenario where the gaps at 48, 86, and 137 au in HD 163296 are opened by a planet at 108 au \protect \citep{dong_etal_2018}. Error bars reflect the combined uncertainties in the CO:H$_2$ conversion factor and the planet mass. The plotted doubling times are minimum values because $\mdotin$ is a maximum accretion rate obtained by assuming all of the mass entering the planet's Bondi or Hill sphere is accreted. The confirmed planets PDS 70b and c have $\mtd \sim t_{\rm age}$. They may be undergoing or may have finished their final mass doublings, just before their parent disc disperses. The interpretation is less clear for the other systems for which $\mtd \ll t_{\rm age}$; a doubling time less than the system age would require us to be observing protoplanets during a special, short-lived period of rapid growth, which seems unlikely. Actual mass-doubling times 
may be longer than the minimum values computed here because actual accretion rates $\dot{M}_{\rm p}$ may be $< \mdotin$; disc gaps may be spatially under-resolved so that $\Sigg$ is overestimated; or disc viscosities are low so that  $M_{\rm p} \sim 10\,\Mearth$, near the low end of the ranges in Table \ref{tab:data}. Lower $\Mp$ increases $\mtd$ because the hypothesised planets would fall into the subthermal regime for which $\mtd \propto 1/\Mp$ (equation \ref{eqn:sum1}). }
  \label{fig:tdouble}
\end{figure*}

Figure \ref{fig:tdouble} compares $\min \, ( t_{\rm double})$ to 
system ages $t_{\rm age}$. A doubling time shorter than the system age
is unlikely as it would require catching the protoplanet during a
short-lived episode of fast growth. We would expect instead
$t_{\rm double} \sim t_{\rm age}$, or $t_{\rm double} > t_{\rm age}$
if the protoplanet has largely finished forming. The protoplanets
PDS 70b and c have $\min \, (t_{\rm double}) \sim t_{\rm age}$; since
$t_{\rm age}$ is comparable to the gas disc's total lifetime, these
objects are either undergoing their last or nearly last doublings, or
have completed their assembly. Unlike the
other entries in Table \ref{tab:data}, PDS 70b and c are detected at a
variety of wavelengths, have astrometry consistent with orbital motion
about their host star, and reside in a large disc cavity.  There
are no confirmed detections among the other putative planets, only a suspicion of existence based on
the observed annular disc gaps they are supposed to have opened (e.g. \citealt{zhang_etal_2018}). Fig.~\ref{fig:tdouble} shows
that for many of these systems, $\min \, (t_{\rm double}) < t_{\rm
  age}$, sometimes by up to 4 orders of
magnitude. There are a number of ways the actual doubling times
$t_{\rm double}$ can exceed our minimum estimates:\footnote{An alternate hypothesis is that the gaps do not actually host planets, but are instead caused by 
local variations in dust grain properties \citep[e.g.][]{birnstiel_etal_2015, hu_etal_2019} or fluid instabilities \citep[e.g.][]{suriano_etal_2018, cui_bai_2021}.} (i) Most obviously in the context of the present work, $\dot{M}_{\rm p} <
\mdotin$; the barriers to permanent accretion of mass from angular momentum and energy may be formidable. \cite{lambrechts_etal_2019b} point out that cooling of the protoplanet's gas envelope may severely limit $\dot{M}_{\rm p}$ (but see \citealt{ginzburg_chiang_2019a} for a simple argument for why cooling is fast once envelope self-gravity becomes important, and also \citealt{kurokawa_tanigawa_2018}). Circumplanetary discs are commonly invoked to remove excess angular momentum, but the mechanism of transport is unknown --- it is not even clear any disc accretes or decretes. Moreover, $\mdotin$ itself may be smaller than we have calculated, if the inflowing material is adiabatic and subsonic \citep{cimerman_etal_2017, fung_etal_2019, moldenhauer_etal_2021, moldenhauer_etal_2022}; 
(ii) Disc gaps may be spatially under-resolved and thus surface densities $\Sigg$ and midplane densities $\rho_{\rm g}$
overestimated; (iii) The non-PDS 70 planets may have masses toward the lower ends of their ranges in Table \ref{tab:data}, closer to $10 M_\oplus$, as would be the case if disc viscosities were low. Lower planet masses would imply longer mass doubling times at subthermal (Bondi) inflow rates.

We plan to leverage our simulations to model the spatial distribution of inflowing material and thereby compute spectral energy distributions. Our preliminary calculations show that much of the accretion power can be re-processed into the mid or far-infrared by circumplanetary dust (see also fig. 6 of \citealt{choksi_chiang_2022}). The protoplanet in HD 163296-G5 (Table \ref{tab:data}) will be targeted by the James Webb Space Telescope later this year \citep{cugno_etal_2023}.

\section*{Acknowledgements}
We thank Chris White for his many hours spent debugging our simulations, and Andrea Antoni and Philipp Kempski for getting us started with $\athena$. \nick{We also thank Aliza Beverage and Isaac Malsky for help with figures, and William Béthune, Yi-Xian Chen, Eve Lee, and
Hidekazu Tanaka for feedback on a draft manuscript.} The anonymous referee provided a thoughtful report that led to substantial improvements in this paper. Simulations were run on the Savio cluster provided by the Berkeley Research Computing program at the University of California, Berkeley, supported by the UC Berkeley Chancellor, Vice Chancellor for Research, and Chief Information Officer. Financial support was provided by NSF AST grant 2205500, and an NSF Graduate Research Fellowship awarded to NC.


\section*{Data availability}
Data and codes are available upon request of the authors.




\bibliographystyle{mnras}
\bibliography{planets_nick} 



\appendix

\section{New data in Table 1}
\label{sec:data}
Table \ref{tab:data} includes entries for AS 209 (G3) and HD 163296 (G4, G5, G234alt) which are not found in the original table of \cite{choksi_chiang_2022}. We describe here the data underlying these new gaps.

\subsection{AS 209-G3}
\cite{zhang_etal_2021} observed a gap in C$^{18}$O emission from $R_{\rm in} = 220$ au to $R_{\rm out} = 260$ au. A thermochemical model fitted to the observed emission yields an H$_2$ surface density of $\Sigg \sim 0.0002\,\rm \gcm$ at the bottom of the gap (their fig. 16). An upper limit on $\Sigg$ can be derived from the  possibility that there is no gap in H$_2$ and that the C$^{18}$O flux is depressed because the CO:H$_2$ abundance ratio is somehow locally depleted beyond what the thermochemical model predicts. This scenario gives $\Sigg \sim 0.1 \,\gcm$ at 240 au (their fig. 5), for an assumed gas-to-dust mass ratio of $10$ (their table 2).  The aspect ratio $\Hp/\Rp = 0.07$ comes from fitting the disc's spectral energy distribution (their table 2). The planet mass of $\Mp = (0.01 - 0.05)\Mj$ is calculated from inserting the normalised gap width $\Delta = (R_{\rm out} - R_{\rm in})/R_{\rm out}$ and a Shakura-Sunyaev viscosity parameter $\alpha = 10^{-5} - 10^{-3}$ into equation 22 of \cite{zhang_etal_2018}.

\subsection{HD 163296-G4, G5, G234alt}
The gap G4 is seen at 137 au in C$^{18}$O and mm continuum emission \citep{isella_etal_2016, teague_etal_2018c, zhang_etal_2021}. The gap G5 is seen at 260 au in near-infrared scattered light \citep{grady_etal_2000}. \cite{isella_etal_2016} fitted the emission from three CO isotopologues assuming an interstellar medium-like CO:H$_2$ ratio and found $\Sigg \sim (0.2-1)\,\rm \gcm$ at 137 au, and $\sim$0.3 $\gcm$ at 260 au (their fig. 2, blue curves; we extrapolated past the edge of their plot, noting that they used the pre-Gaia source distance of 122 pc which is about 20\% too large). These surface densities are similar to values derived from the thermochemical models of \citet[][their fig. 16]{zhang_etal_2021}, $\Sigg \sim 0.3 \, \gcm$ and $\sim$$0.1 \,\gcm$, respectively. If instead the CO:H$_2$ abundance ratio is lower than predicted by the latter models and the gas-to-dust ratio is 60 (their table 2), then $\Sigg \sim 7\,\gcm$ at 137 au and $\Sigg \sim 1.5 \,\gcm$ at 260 au (their figure 5). In Table \ref{tab:data} we summarise these results as $\Sigg = (0.2 - 7)\,\gcm$ for G4 and $(0.1 - 1.5)\,\gcm$ for G5. The local aspect ratio $\Hp/\Rp = 0.09$ comes from a fit to the spectral energy distribution (table 2 of \citealt{zhang_etal_2021}).

For G4, \citet[][their table 5]{zhang_etal_2021} estimated $\Mp \sim 0.005 \Mj$ using the width of the mm continuum gap and an assumed $\alpha = 10^{-4}$. In our paper we entertain $\alpha$ as small as $10^{-5}$ and therefore obtain a lower limit on $\Mp$ of $\sim$$0.002 \Mj$ using the empirical scaling relation $\Mp \propto \alpha^{1/3}$ \citep{zhang_etal_2018}. 
Gap G5 is only observed in near-infrared scattered light \citep{grady_etal_2000}. Assuming the grains most visible at these wavelengths trace the gas, we use equation 22 of \cite{zhang_etal_2018}, which is calibrated for gas gaps, to infer a minimum $\Mp \sim 0.01 \Mj$ (based on the width of the scattered-light gap of 40 au and $\alpha = 10^{-5}$). Upper limits on $\Mp$ of 1.3$\Mj$ \citep[G4;][]{teague_etal_2018c} and 2$\Mj$ \citep[G5;][]{pinte_etal_2018} are derived from examining non-Keplerian velocities.

\cite{dong_etal_2018} showed that the gaps G2, G3, and G4 in HD 163296 do not need to host  planets. Instead, a single planet with $\Mp \sim 0.2\,\Mj$ orbiting at 108 au can open all three gaps if the disc has low enough viscosity (their fig.~9). The entry ``HD 163296-G234alt'' in Table \ref{tab:data} refers to this scenario. The range of surface densities $\Sigg$ near 108 au are drawn from figs.~5 and 16 of \cite{zhang_etal_2021}.

\section{Convergence Tests}
\label{sec:convergence}
\subsection{Grid resolution}
\begin{figure} 
\includegraphics[width=\columnwidth]{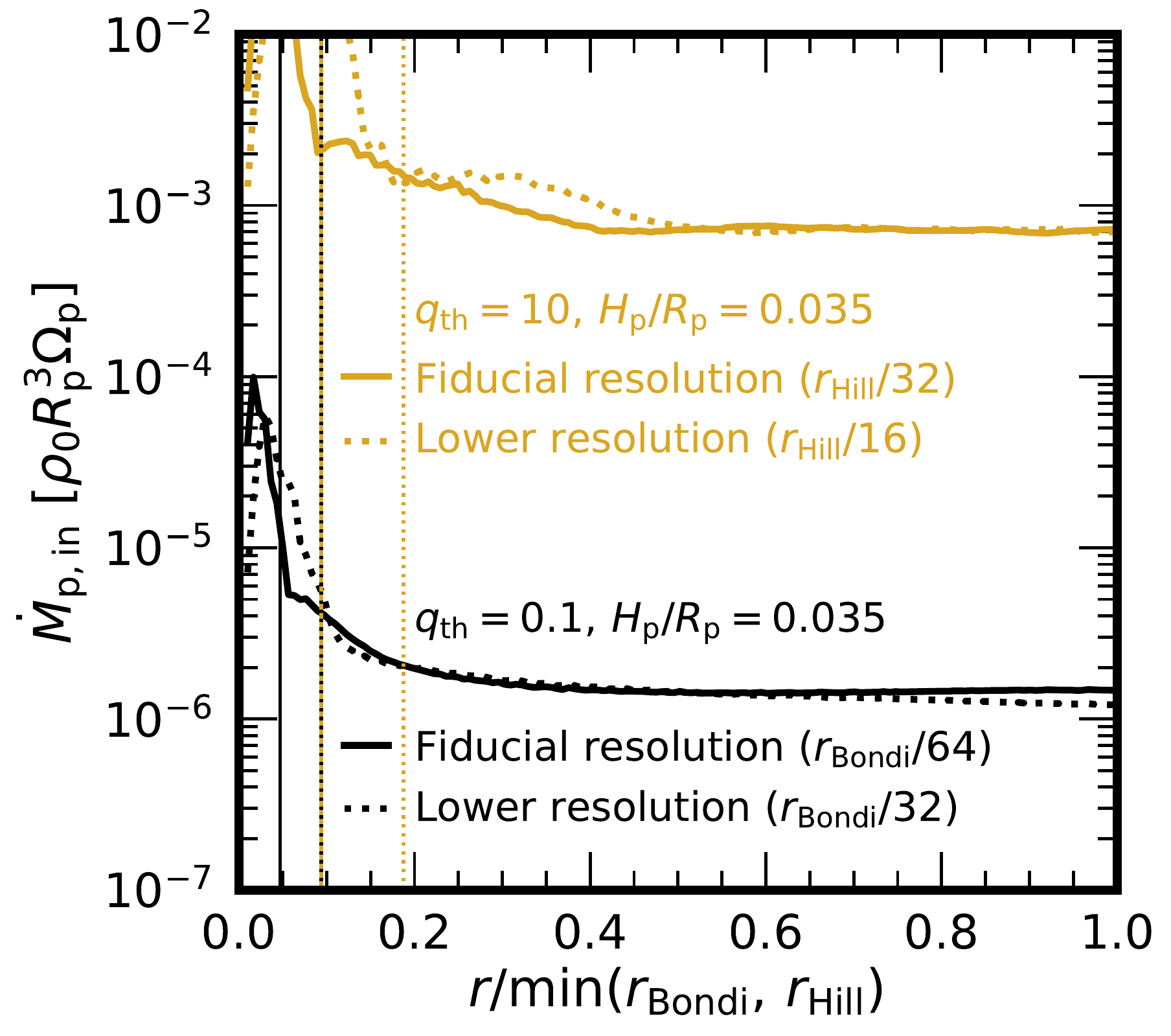}
\caption{Comparison of time-averaged inflow rates $\mdotin$ obtained at standard resolution (solid curves) and $2\times$ 
lower resolution (dotted curves) for $\qth = 0.1$ (black) and $\qth = 10$ (gold). Minimum grid cell widths are shown in parentheses. Vertical lines mark gravitational potential softening radii, equal to three times the minimum cell width.
Inflow rates $\mdotin$ from $0.2 \lesssim r/\rb \lesssim 1$ for $\qth = 0.1$, and $0.4 \lesssim r/\rh \lesssim 1$ for $\qth = 10$, appear to have largely converged with resolution.}
\label{fig:convergence}
\end{figure}

We re-ran some of our \rev{non-sink-cell} $\athena$ simulations at lower resolution. Compared to our fiducial setup, these runs had grid cell widths that were twice as large. As Figure \ref{fig:convergence} shows, changing the resolution over this range does not affect our finding that $\mdotin$ is nearly constant from $0.2 \lesssim r/\rb \lesssim 1$ for subthermal planets, and $0.4 \lesssim r/\rh \lesssim 1$ for superthermal planets. In these regions $\mdotin$ differs by at most tens of percent between the two resolutions.

\begin{figure} 
\includegraphics[width=\columnwidth]{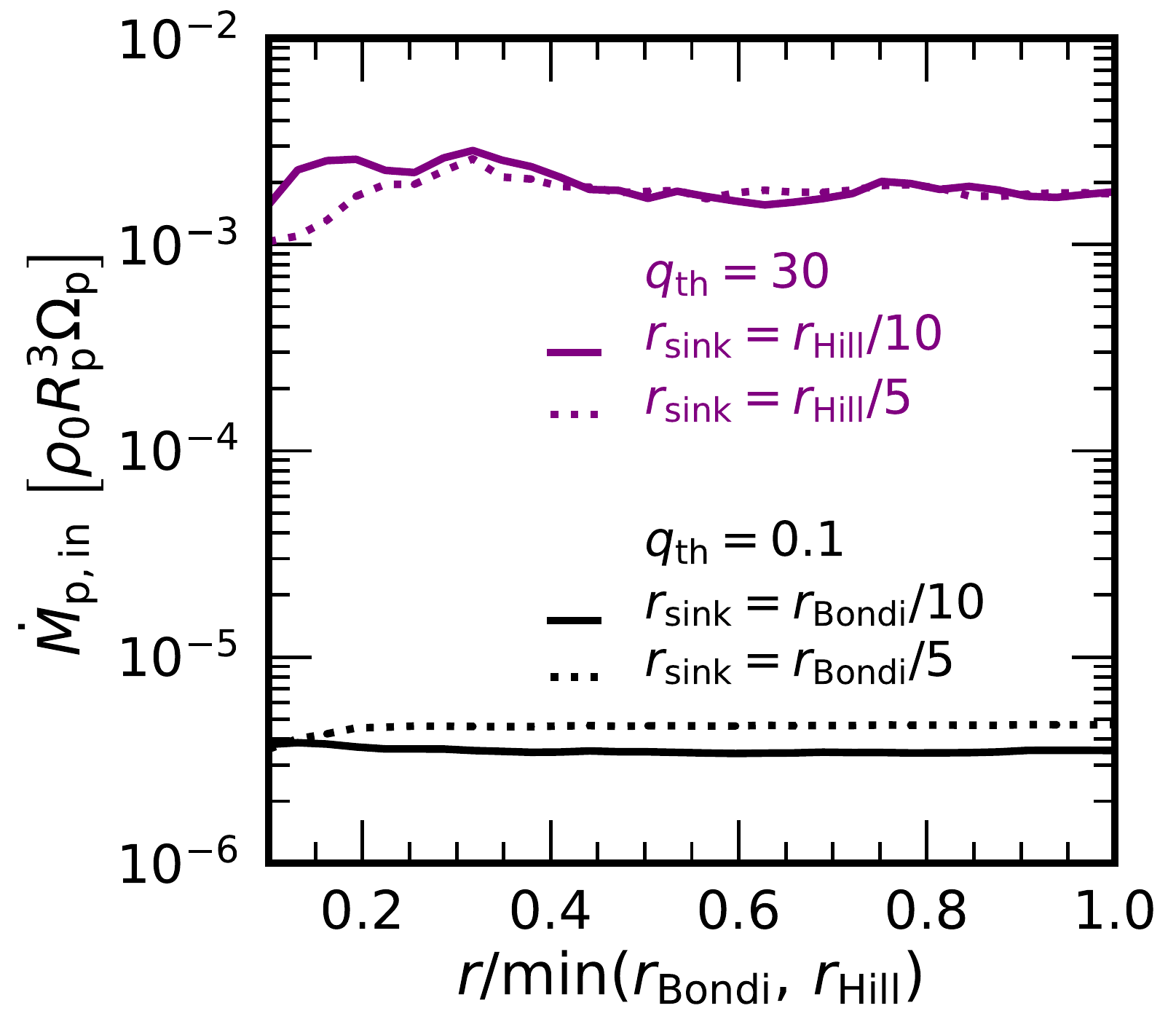}
\caption{ \rev{Comparison of time-averaged inflow rates $\mdotin$ obtained using $r_{\rm sink} = \minrh/10$ (solid) and an alternate $r_{\rm sink} = \minrh/5$ (dotted). Inflow rates have converged with $r_{\rm sink}$ to within 20\% for $\qth = 0.1$ (black) and to within a few percent for $\qth = 30$ (purple). }
}
\label{fig:rsink}
\end{figure}

\rev{
\subsection{Sink-cell domain size}
Our fiducial simulations including sink cells depleted the gas density interior to $r_{\rm sink} = \minrh/10$. Figure \ref{fig:rsink} shows that increasing $r_{\rm sink}$ to $\minrh/5$ does not affect our finding of a nearly constant $\mdotin$ between $r_{\rm sink}$ and $\minrh$. Inflow rates between the two $r_{\rm sink}$ models differ by 20\% for $\qth = 0.1$ and by a few percent for $\qth = 30$.
}

\bsp	
\label{lastpage}
\end{document}